\documentclass[twocolumn]{aastex7}

\usepackage[T1]{fontenc}          
\usepackage{lmodern}              
\usepackage{textcomp}             
\usepackage{microtype}

\usepackage{natbib}
\usepackage{amsmath}
\usepackage{multirow}
\usepackage[normalem]{ulem}
\usepackage{xcolor}
\usepackage{xspace}
\usepackage{enumitem}   
\usepackage{graphicx}
\usepackage{wrapfig}
\usepackage{lineno}
\usepackage{comment}

\newcommand{\etaearth}{$\eta_{\oplus}$\xspace}
\newcommand{\gammaearth}{$\Gamma_{\oplus}$}
\newcommand{\Rearth}{$R_\oplus$\xspace}
\newcommand{\Mearth}{$M_\oplus$\xspace}
\newcommand{\Searth}{$S_\oplus$\xspace}

\newcommand{\Msun}{$M_\odot$\xspace}
\newcommand{\gaia}{\texttt{Gaia}\xspace}

\newcommand{\kepler}{\emph{Kepler}\xspace}

\newcommand{\ktwo}{\emph{K2}\xspace}

\newcommand{\Mjup}{$M_\mathrm{Jup}$\xspace}
\let\oldmaketitle\maketitle
\renewcommand{\maketitle}{\oldmaketitle\setcounter{footnote}{0}}

\submitjournal{PASP}
\received{July 2025}
\revised{November 2025}
\definecolor{tropicalrainforest}{rgb}{0.0, 0.46, 0.37}
\definecolor{plum}{HTML}{88498f}
\definecolor{deeplilac}{rgb}{0.6, 0.33, 0.73}
\definecolor{brightube}{rgb}{0.82, 0.62, 0.91}
\definecolor{midnightblue}{rgb}{0.1, 0.1, 0.44}
\definecolor{michelegreen}{rgb}{0, 0.7, 0}
\definecolor{mahogany}{HTML}{670A0A}
\definecolor{sakheenavy}{HTML}{57799F}
\definecolor{jamiered}{HTML}{F26035}

\begin{document}
\title{Are We There Yet? Challenges in Quantifying the Frequency of Earth Analogs in the Habitable Zone}

\correspondingauthor{Rachel B. Fernandes}
\email{rbf5378@psu.edu}

\collaboration{all}{Lead Authors\footnote{All lead authors have contributed equally to this work}}
\author[0000-0002-3853-7327]{Rachel B. Fernandes}
\altaffiliation{President's Postdoctoral Fellow}
\affil{Department of Astronomy \& Astrophysics, 525 Davey Laboratory, The Pennsylvania State University, University Park, PA 16802, USA}
\affil{Center for Exoplanets and Habitable Worlds, 525 Davey Laboratory, The Pennsylvania State University, University Park, PA 16802, USA}
\email{rbf5378@psu.edu}

\author[0000-0001-9397-4768]{Samson Johnson}
\altaffiliation{NASA Postdoctoral Program Fellow}
\affil{Jet Propulsion Laboratory, California Institute of Technology, 4800 Oak Grove Drive, Pasadena, CA 91109, USA}
\email{johnson.7080@osu.edu}

\author[0000-0003-4500-8850]{Galen J. Bergsten}
\affil{Lunar and Planetary Laboratory, The University of Arizona, Tucson, AZ 85721, USA}
\email{gbergsten@arizona.edu}

\author[0000-0002-6673-8206]{Sakhee Bhure}
\affil{Centre for Astrophysics, University of Southern Queensland, Toowoomba, QLD 4350, Australia}
\email{Sakhee.Bhure@unisq.edu.au}

\author[0000-0001-8153-639X]{Kiersten M. Boley}
\altaffiliation{NASA Sagan Fellow}
\affil{Earth \& Planets Laboratory, Carnegie Institution for Science, Washington, DC, 20015, USA}
\email{kboley@carnegiescience.edu}

\author[0000-0001-7119-1105]{Alan P. Boss}
\affil{Earth \& Planets Laboratory, Carnegie Institution for Science, Washington, DC, 20015, USA}
\email{aboss@carnegiescience.edu}

\author[0000-0003-0081-1797]{Steve Bryson}
\affil{NASA Ames Research Center, Moffett Field, CA 94035, USA}
\email{steve.bryson@nasa.gov}

\author[0000-0003-1827-9399]{William DeRocco}
\affil{Maryland Center for Fundamental Physics, University of Maryland, College Park, 4296 Stadium Drive, College Park, MD 20742, USA}
\affil{Department of Physics \& Astronomy, The Johns Hopkins University, 3400 N. Charles Street, Baltimore, MD 21218, USA}
\email{derocco@umd.edu}

\author[0000-0001-6320-7410]{Jamie Dietrich}
\affil{School of Earth and Space Exploration, Arizona State University, 781 Terrace Mall, Tempe, AZ 85287}
\email{jdietri3@asu.edu}

\author[0000-0002-4531-6899]{Alison Duck}
\affil{Department of Astronomy, The Ohio State University, Columbus, OH 43210, USA}
\email{duck.18@buckeyemail.osu.edu}

\author[0000-0002-8965-3969]{Steven Giacalone}
\altaffiliation{NSF Astronomy and Astrophysics Postdoctoral Fellow}
\affil{Department of Astronomy, California Institute of Technology, Pasadena, CA 91125, USA}
\email{giacalone@astro.caltech.edu}

\author[0000-0002-5463-9980]{Arvind F.\ Gupta}
\affil{U.S. National Science Foundation National Optical-Infrared Astronomy Research Laboratory, 950 N.\ Cherry Ave., Tucson, AZ 85719, USA}
\email{arvind.gupta@noirlab.edu}

\author[0000-0002-5223-7945]{Matthias Y. He}
\altaffiliation{NASA Postdoctoral Program Fellow}
\affil{NASA Ames Research Center, Moffett Field, CA 94035, USA}
\email{heymatthias94@gmail.com}

\author[0000-0001-9269-8060]{Michelle Kunimoto}
\affil{Department of Physics and Astronomy, University of British Columbia, 6224 Agricultural Road, Vancouver, BC V6T 1Z1, Canada}
\email{mkuni@phas.ubc.ca}

\author[0000-0001-5847-9147]{Kristo Ment}
\affil{Department of Astronomy \& Astrophysics, 525 Davey Laboratory, The Pennsylvania State University, University Park, PA 16802, USA}
\affil{Center for Exoplanets and Habitable Worlds, 525 Davey Laboratory, The Pennsylvania State University, University Park, PA 16802, USA}
\email{kxm821@psu.edu}

\author[0000-0002-3022-6858]{Sheila Sagear}
\affil{Department of Astronomy, University of Florida, 211 Bryant Space Science Center, Gainesville, FL 32611}
\email{ssagear@ufl.edu}

\author[0000-0003-2565-7909]{Michele L. Silverstein}
\altaffiliation{NRC Research Associate}
\affil{Naval Research Laboratory, 4555 Overlook Avenue SW, Washington, DC 20375, USA}
\email{mlsilverstein@proton.me}

\author[0000-0001-6873-8501]{Kendall Sullivan}
\affil{Department of Astronomy and Astrophysics, University of California Santa Cruz, Santa Cruz, CA 95064, USA}
\email{ksulliv4@ucsc.edu}

\author[0000-0002-1864-6120]{Eliot Halley Vrijmoet}
\affil{Five College Astronomy Department, Smith College, Northampton, MA 01063, USA}
\affil{RECONS Institute, Chambersburg, PA 17201, USA}
\email{evrijmoet@smith.edu}

\author[0000-0002-4309-6343]{Kevin Wagner}
\affil{Department of Astronomy, University of Arizona, 933 N Cherry Ave, Tucson, AZ 85719}
\email{kevinwagner@arizona.edu}

\author[0000-0002-4235-6369]{Robert F. Wilson}
\affil{Department of Astronomy, University of Maryland, College Park, MD 20742, USA}
\affil{NASA Goddard Space Flight Center, Greenbelt, MD, 20771, USA}
\email{robert.f.wilson@nasa.gov}

\collaboration{all}{Contributing Authors}
\author[0000-0001-9256-5508]{Lucas Brefka}
\affil{Department of Astronomy \& Astrophysics, 525 Davey Laboratory, The Pennsylvania State University, University Park, PA 16802, USA}
\affil{Center for Exoplanets and Habitable Worlds, 525 Davey Laboratory, The Pennsylvania State University, University Park, PA 16802, USA}
\email{lfb5454@psu.edu}

\author[0000-0002-4951-8025]{Ruslan Belikov}
\affil{NASA Ames Research Center, Moffett Field, CA 94035, USA}
\email{ruslan.belikov-1@nasa.gov}

\author[0000-0001-6703-0798]{Aritra Chakrabarty}
\affil{NASA Ames Research Center, Moffett Field, CA 94035, USA}
\altaffiliation{NASA Postdoctoral Program Fellow}
\email{aritra.astrophysics@gmail.com}

\author[0000-0002-8035-4778]{Jessie L.\ Christiansen}
\affil{NASA Exoplanet Science Institute, IPAC, California Institute of Technology, Pasadena, CA 91125 USA}
\email{christia@ipac.caltech.edu}

\author[0000-0002-5741-3047]{David R. Ciardi}
\affil{NASA Exoplanet Science Institute, IPAC, California Institute of Technology, Pasadena, CA 91125 USA}
\email{ciardi@ipac.caltech.edu}

\author[0000-0002-1092-2995]{Anne Dattilo}
\affil{Department of Astronomy \& Astrophysics, University of California Santa Cruz, 1156 High Street, Santa Cruz, CA, 95064, USA}
\email{adattilo@ucsc.edu}

\author[0000-0003-0199-9699]{Evan Fitzmaurice}
\affil{Department of Astronomy \& Astrophysics, 525 Davey Laboratory, The Pennsylvania State University, University Park, PA 16802, USA}
\affil{Center for Exoplanets and Habitable Worlds, 525 Davey Laboratory, The Pennsylvania State University, University Park, PA 16802, USA}
\affil{Institute for Computational and Data Sciences, The Pennsylvania State University, University Park, PA 16802, USA}
\altaffiliation{Institute for Computational and Data Sciences Scholar}
\email{exf5296@psu.edu}

\author[0000-0001-6545-639X]{Eric B.\ Ford}
\affil{Department of Astronomy \& Astrophysics, 525 Davey Laboratory, The Pennsylvania State University, University Park, PA 16802, USA}
\affil{Center for Exoplanets and Habitable Worlds, 525 Davey Laboratory, The Pennsylvania State University, University Park, PA 16802, USA}
\affil{Institute for Computational and Data Sciences,  Penn State University, University Park, PA, 16802, USA}
\affil{Center for Astrostatistics,  525 Davey Laboratory, 251 Pollock Road, University Park, PA, 16802, USA}
\email{eford@psu.edu}

\author[0009-0000-1825-4306]{Andrew Hotnisky}
\affil{Department of Astronomy \& Astrophysics, 525 Davey Laboratory, The Pennsylvania State University, University Park, PA 16802, USA}
\affil{Center for Exoplanets and Habitable Worlds, 525 Davey Laboratory, The Pennsylvania State University, University Park, PA 16802, USA}
\email{amh7996@psu.edu}

\author[0000-0002-7227-2334]{Sinclaire Jones}
\affil{Department of Astronomy, The Ohio State University, Columbus, OH 43210, USA}
\email{jones.7887@buckeyemail.osu.edu}

\author[0000-0002-9811-5521]{Aman Kar}
\affil{Department of Physics and Astronomy, Georgia State University, Atlanta, GA 30303, USA}
\affil{RECONS Institute, Chambersburg, PA 17201, USA}
\email{akar5@gsu.edu}

\author[0000-0002-5893-2471]{Ravi Kopparapu}
\affil{NASA Goddard Space Flight Center, Greenbelt, MD, 20771, USA}
\email{ravikumar.kopparapu@nasa.gov}

\author[0000-0001-6508-5736]{Nataliea Lowson}
\affil{Department of Physics and Astronomy, University of Delaware, 217 Sharp Lab, Newark, DE 19716, USA}
\altaffiliation{Annie Jump-Cannon Fellow}
\email{nlowson@udel.edu}

\author[0000-0003-2008-1488]{Eric E. Mamajek}
\affil{Jet Propulsion Laboratory, California Institute of Technology, 4800 Oak Grove Drive, Pasadena, CA 91109, USA}
\affil{Department of Physics and Astronomy, University of Rochester, Rochester, NY 14627-0171, USA}
\email{eric.mamajek@jpl.nasa.gov}

\author[0000-0003-4205-4800]{Bertrand Mennesson}
\affil{Jet Propulsion Laboratory, California Institute of Technology, 4800 Oak Grove Drive, Pasadena, CA 91109, USA}
\email{bertrand.mennesson@jpl.nasa.gov}

\author[0000-0003-1227-3084]{Michael R. Meyer}
\affil{Department of Astronomy, The University of Michigan, 1085 S. University Ave, 
Ann Arbor, MI 48109}
\email{mrmeyer@umich.edu}

\author[0000-0003-3130-2282]{Sarah Millholland}
\affil{Department of Physics, Massachusetts Institute of Technology, Cambridge, MA 02139, USA
MIT Kavli Institute for Astrophysics and Space Research, Massachusetts Institute of Technology, Cambridge, MA 02139, USA}
\email{sarah.millholland@mit.edu}

\author[0000-0002-1078-9493]{Gijs D. Mulders}
\affil{Instituto de Astrof\'isica, Pontificia Universidad Cat\'olica de Chile, Av. Vicu\~na Mackenna 4860, 7820436 Macul, Santiago, Chile} 
\email{gdmulders@gmail.com}

\author[0000-0001-7106-4683]{Susan E. Mullally}
\affil{Space Telescope Science Institute, 3700 San Martin Drive, Baltimore, MD 21218, USA}
\email{smullally@stsci.edu}

\author[0009-0004-1245-092X]{Arjun Murlidhar}
\affil{Department of Astronomy, The Ohio State University, Columbus, OH 43210, USA}
\email{murlidhar.4@buckeyemail.osu.edu}

\author[0000-0001-7962-1683]{Ilaria Pascucci}
\affil{Lunar and Planetary Laboratory, The University of Arizona, Tucson, AZ 85721, USA}
\email{pascucci@arizona.edu}

\author[0000-0003-1080-9770]{Darin Ragozzine}
\affil{Brigham Young University, Department of Physics and Astronomy, N283 ESC, Provo, UT 84602, USA}
\email{darin_ragozzine@byu.edu}

\author[0000-0003-0149-9678]{Paul Robertson}
\affil{Department of Physics \& Astronomy, The University of California, Irvine, Irvine, CA 92697, USA}
\email{probert1@uci.edu}

\author[0000-0002-2805-7338]{Karl Stapelfeldt}
\affil{Jet Propulsion Laboratory, California Institute of Technology, 4800 Oak Grove Drive, Pasadena, CA 91109, USA}
\email{karl.r.stapelfeldt@jpl.nasa.gov}

\author[0000-0001-6160-5888]{Jason Wright}
\affil{Department of Astronomy \& Astrophysics, 525 Davey Laboratory, The Pennsylvania State University, University Park, PA 16802, USA}
\affil{Center for Exoplanets and Habitable Worlds, 525 Davey Laboratory, The Pennsylvania State University, University Park, PA 16802, USA}
\affil{Penn State Extraterrestrial Intelligence Center, 525 Davey Laboratory, 251 Pollock Road, Penn State, University Park, PA, 16802, USA}
\email{astrowright@gmail.com}

\begin{abstract}

Searching for life elsewhere in the universe is one of the most highly prioritized pursuits in astronomy today. However, the ability to observe evidence of Earth-like life through biosignatures is limited by the number of planets in the solar neighborhood with conditions similar to Earth. The occurrence rate of Earth-like planets in the habitable zones of Sun-like stars, \etaearth, is therefore crucial for addressing the apparent lack of consensus on its value in the literature. Here we present a review of the current understanding of \etaearth. We first provide definitions for parameters that contribute to \etaearth. Then, we discuss the previous and current estimated parameter values and the context of the limitations on the analyses that produced these estimates. We compile an extensive list of the factors that go into any calculation of \etaearth, and how detection techniques and surveys differ in their sensitivity and ability to accurately constrain \etaearth. Understanding and refining the value of \etaearth is crucial for upcoming missions and telescopes, such as the planned Habitable Worlds Observatory and the Large Interferometer for Exoplanets, which aim to search for biosignatures on exoplanets in the solar neighborhood.
\end{abstract}


\section{Introduction} \label{sec:intro}
Identifying an Earth analog --- broadly, a terrestrial planet orbiting a Sun-like star that is capable of harboring life --- has been a long-standing goal of the exoplanet community. The Decadal Survey on Astronomy and Astrophysics 2020 (Astro2020) listed the discovery of such worlds as well as subsequent searches for biosignatures of life as one of the highest priorities in the field \citep{Decadal}. However, this task is challenging in part because of the weak constraints (illustrated below) on how common potentially habitable planets are, limiting our ability to design optimized surveys in search of such worlds. Estimated occurrence rates of potentially habitable planets are relied upon to calculate mission yield predictions and can therefore affect the scope and design decisions of upcoming mission concepts such as the Habitable Worlds Observatory \footnote{\url{https://science.nasa.gov/astrophysics/programs/habitable-worlds-observatory/}} (HWO; as recommended by Astro2020), the Large Interferometer for Exoplanets\footnote{\url{https://life-space-mission.com/}} (LIFE), extremely large telescopes, and other facilities. 

The frequency of Earth-like (defined in the next paragraph) planets in the habitable zones (HZ) of Sun-like (typically, FGK spectral type dwarf) stars has been loosely referred to as \etaearth (``Eta-Earth''). The term \etaearth is closely related to (and was likely inspired by) the Drake equation, which includes a similar term $n_e$ representing ``the mean number of planets in each planetary system with environments permitting the development of life'' \citep{Drake1965}. Indeed, in popular presentations the connection between \etaearth and the Drake equation is often made explicitly, either to motivate the measurement of \etaearth or to explain the significance of that measurement. A measurement of \etaearth (and its equivalent for M dwarfs) thus moves us towards a measurement of $n_e$, which would be made more precise once the fraction of such planets with ``environments permitting the development of life'' is better known. However, the exact parameters and definitions relevant to calculating \etaearth (e.g., what constitutes an Earth-like planet, a Sun-like star, or the boundaries of a habitable zone) have evolved over time due to changes in our understanding of the conditions that support planetary habitability.

\begin{figure*}[!htpb]
    \centering    \includegraphics[width=1.0\linewidth]{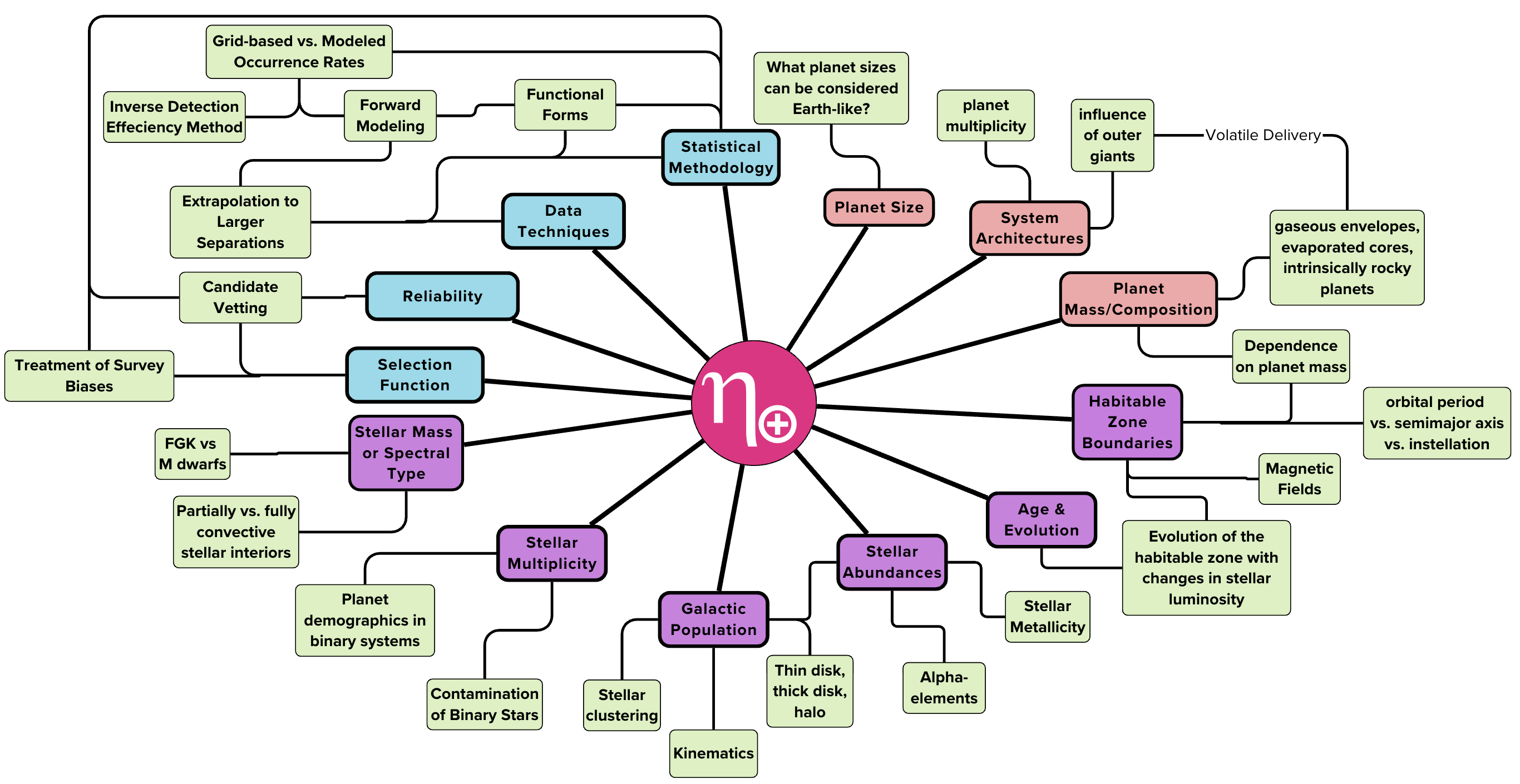}
    \caption{Depiction of the many factors affecting \etaearth and measurements therein. The three main colors offer a loose categorization into planetary, stellar, and statistical issues (red, purple, and blue, respectively), with subcategories (in green) branching out and interconnecting}.
    \label{fig:etaEarthOverview}
\end{figure*}

For the purposes of measuring \etaearth, a planet has been typically considered ``Earth-like'' if it is thought to have a primarily rocky composition and is similar to Earth in size. Planets that are too small are unlikely to be able to retain an atmosphere or exhibit plate tectonics \citep{Kasting1993}. The lower mass limit for planetary habitability is thought to be around 0.3 \Mearth~\citep{Raymond2007}, which corresponds to an estimated radius of 0.72 \Rearth~\citep[][based on the empirical mass-radius relationship by \citealt{chen2017prob}]{Zink2019multiplicity}. Studies have employed various radii within the range of 0.5--1 \Rearth~for the lower limit of a potentially habitable planet. On the other hand, planets substantially larger than Earth tend to be enshrouded in significant volatile-rich envelopes; classifying them as sub-Neptunes. Consequently, the terms ``Earth-like'' and ``Earth-sized'' are often used interchangeably, because an Earth-sized planet in the HZ is expected to have a primarily rocky (and therefore, Earth-like) composition. The transition between the super-Earth (mostly rocky planets larger than Earth) and sub-Neptune regimes is expected to occur between 1.5--2 \Rearth~\citep{Weiss2014,rogers2015most}, evidenced by a corresponding gap in the radius distribution of small planets \citep{fulton2017california, hardegree2020scaling}. Subsequently, studies have frequently chosen an upper boundary between 1.5--2.0 \Rearth~for the radius of a potentially habitable planet.

The HZ has been defined as the orbital region around the star where estimated surface temperatures would permit liquid water on the surface of a terrestrial planet with an Earth-like atmosphere at the present day\footnote{A more restrictive definition would be obtained by using the Continuously Habitable Zone \citep{Hart1979}, which requires the planet to have remained habitable throughout its evolution history. However, reliably estimating planet evolution over time is in itself a complicated problem.}. This region is typically quantified by a range of orbital periods, semi-major axes, or surface instellation (the amount of stellar radiation power received per surface area) values. \citet{Kasting1993} used a one-dimensional climate model to produce \textit{conservative} estimates of 0.95 au and 1.37 au for the inner and outer edges of the HZ in the present day Solar System, respectively. As explained by \citet{Kasting1993}, interior to 0.95 au, photolysis and hydrogen escape are expected to lead to permanent water loss, whereas beyond 1.37 au, the formation of CO$_2$ clouds would cool the planet's surface by increasing its albedo and lowering the convective lapse rate. They noted that the boundaries of the HZ are expected to evolve outwards as the star ages and becomes more luminous. The width and location of the HZ also depend on stellar luminosity, with less luminous stars possessing HZs that are smaller in size and occur closer in \citep[e.g.,]{Kopparapu2013}. A later study showed that CO$_2$ clouds generally warm a planet's climate instead \citep{Forget1997}. \citet{Kopparapu2013} used an updated one-dimensional radiative-convective, cloud-free climate model to revise the semi-major axis range of the current Solar System conservative HZ to 0.99--1.70 au, with the outer HZ boundary defined by the maximum greenhouse limit and the effect of CO$_2$ clouds neglected completely. Alternatively, \citet{petigura2013prevalence} proposed a simpler definition of the HZ based on the analysis of \kepler mission data. They defined the HZ as the instellation range 0.25--4 \Searth~(that is, within a factor of 4 of the current Earth), corresponding to a semi-major axis range of 0.5--2 au in the current Solar System. Studies \citep[etc.]{Kasting1993,Kopparapu2013} have also considered more \textit{optimistic} estimates for HZ boundaries, where the inner edge is typically calculated under the assumption that Venus had liquid water on its surface early in its history (but subsequently lost it), and the outer edge is calculated based on the inference that early Mars had liquid water on its surface. Overall, as shown in Section \ref{sec:etaEarthHistory}, studies estimating \etaearth have used various different HZ boundaries over the years.

The Exoplanet Task Force was established at the request of NSF and NASA in the late 2000s with the explicit charge of recommending a 15-year strategy to detect Earth-like planets and study their habitability. Its official report \citep{Lunine2008} defined \etaearth as ``the fraction of stars\footnote{Modern studies of \etaearth typically estimate the total number of potentially habitable planets per star, whereas the Exoplanet Task Force defined \etaearth as the fraction of stars ($\leq 1$) with at least one potentially habitable planet. For consistency and clarity, we adopt decimal fractions for \etaearth to mean ``planets per star" and percentage values for ``fraction of stars with planets".} that have at least one potentially habitable planet''. The Task Force interpreted a potentially habitable planet as one that is close to the size of the Earth (between 0.5--2 \Rearth, or between 0.1--10 \Mearth~in mass) and that orbits within the HZ of the star. Based on the handful of Earth-like exoplanet candidates available at the time, \citet{Vogt2010} estimated a lower bound on \etaearth of about 15\%.

In March 2009, NASA launched the \kepler space telescope, which was a planet-hunting mission designed to detect Earth-sized planets orbiting Sun-like stars in a predefined portion of the galaxy. Discoveries from the \kepler mission and its successor, the \ktwo mission, catalyzed many planet demographics studies in the decade that followed. After a number of these studies endeavored to measure \etaearth and found a large dispersion in the results \citep[e.g.,][etc.]{petigura2013prevalence, ForemanMackey2014, Silburt2015, Burke2015, Mulders2018}, NASA's Exoplanet Exploration Program formed Study Analysis Group 13 (SAG13) to provide a consolidated value for \etaearth in preparation for the Astro2020 Decadal Survey and to address the need to estimate the expected yield of habitable worlds that might be detected and characterized by various space mission concepts. The SAG13 final report was released in 2017\footnote{Available from \url{https://exoplanets.nasa.gov/exep/exopag/sag}} and concluded that the value for \etaearth was close to 0.6, higher than previous estimates but slightly lower than the \kepler team's preliminary result of 0.77 for G and K dwarfs \citep{Burke2015}.\footnote{However, it should be noted \citet{Burke2015} used a somewhat different habitable planet definition compared to that used by the Exoplanet Task Force Report.} Subsequently, the final \kepler team result \citep{Bryson2021} used the updated \kepler DR25 planet candidate catalog and \gaia-based stellar properties to estimate a conservative HZ occurrence rate between 0.37--0.60 and an optimistic HZ occurrence rate between 0.58--0.88. To standardize the limits on planetary radius for the purposes of calculating \etaearth, the SAG13 final report also recommended that studies provide separate values for \etaearth~using planetary radii between 0.5--1.5 \Rearth~as well as 1.0--1.5 \Rearth. Post-SAG13 studies have continued to produce discrepant estimates for \etaearth, which is demonstrated in Section \ref{sec:etaEarthHistory} of this work.

It should be emphasized that not all Earth-sized planets in the HZ are necessarily habitable, and not all habitable planets must reside in the HZ of their host stars. There are many factors that must work in conjunction, all contributing to the habitability of a planet. These properties may include the water content of the planet \citep{Cowan2014,Kurosaki2014}, the role of plate tectonics \citep{Noack2014} and planetary magnetic fields \citep{Airapetian2017,Tilley2019}, the core mass fraction \citep{Noack2017}, etc. We restrict our discussion to some fairly basic properties of exoplanets that can be explicitly measured with current technology. A broader depiction of the many factors affecting \etaearth is displayed in Figure \ref{fig:etaEarthOverview}. We also acknowledge that alternative parameterizations of the frequency of habitable planets have appeared in the literature, such as \gammaearth~\citep[``Gamma-Earth'';][]{ForemanMackey2014}, which is an occurrence rate density that is designed to be independent of one's choice of habitable planet size or HZ boundaries\footnote{For a review of \gammaearth~estimates in the literature, see \citet{Kunimoto2020}.}. This report focuses explicitly on \etaearth as its value is directly used in developing missions to find and characterize Earth-like planets; however, many of the factors outlined here apply to parameters such as \gammaearth~as well.

Given the rapid progress in the field since the release of the SAG13 report and the prioritization of the Habitable Worlds Observatory by Astro2020, finding pathways to refining \etaearth has become a priority for the exoplanet demographics community. As a first step in this direction, this work aims to review some of the major efforts to measure \etaearth, discuss attributes of planetary systems not being accounted for in determining \etaearth, and describe how different exoplanet detection techniques can contribute to improving our understanding of \etaearth.

First, we detail the chronological changes in the estimated value of \etaearth and discuss the causes of these changes in Section \ref{sec:etaEarthHistory}. In Section \ref{sec:missingAspects}, we explore a number of aspects of planetary systems that are currently not being fully considered in computing \etaearth that could impact its value. We then present a summary of the future prospects of improving \etaearth estimates from current and future surveys and missions in Section \ref{sec:futureProspects}. Finally, Section \ref{sec:conclusions} provides the summary and conclusions of our analysis. We would like to note that the scope of this work does not include providing a new estimate for \etaearth. Rather, we aim to provide a consolidated summary of the past and present state of the field as well as future directions, which may serve as a helpful guide to any upcoming studies aiming to calculate \etaearth.


\section{Chronology and Challenges for Previous \etaearth Estimates}
\label{sec:etaEarthHistory}

\begin{figure*}[!htpb]
\begin{center}
\includegraphics[width=0.98\textwidth]{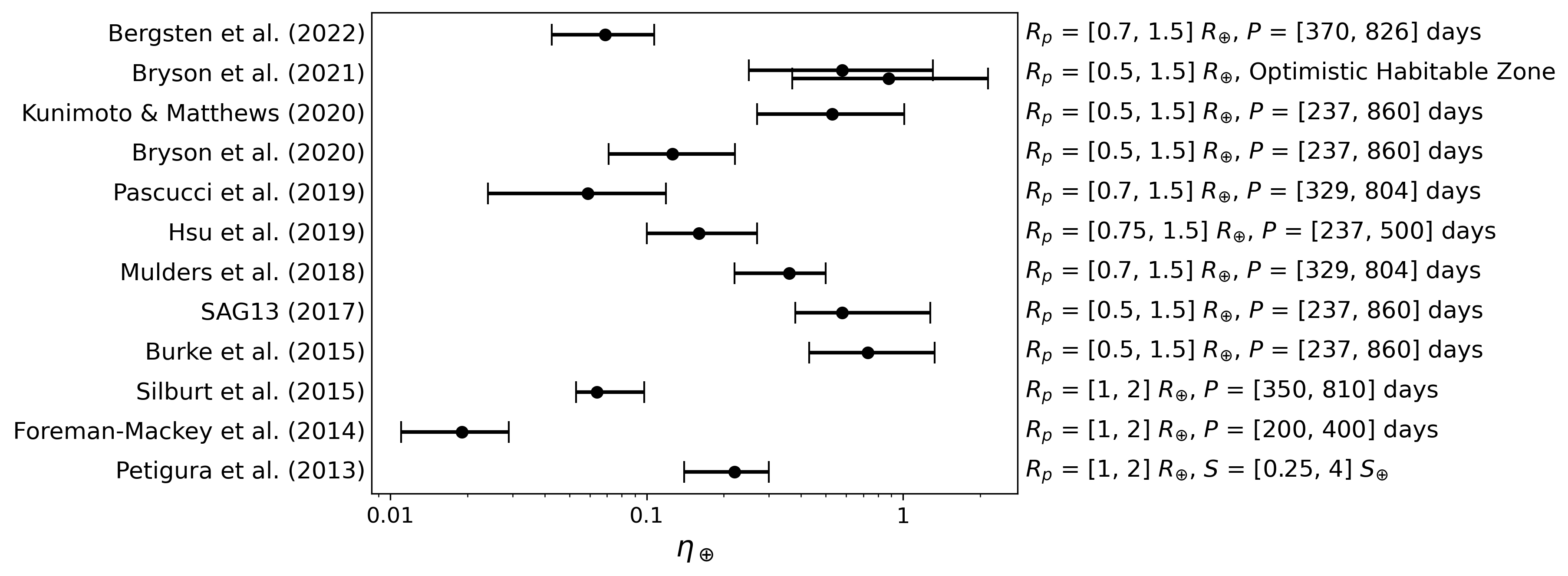} 
\caption{{A comparison of previous \etaearth estimates in log-scale, with the definition of the \etaearth regime shown on the right. \cite{Bryson2021} integrated over the optimistic habitable zone \citep{Kopparapu2013} based on individual stars' stellar effective temperature ($237 - 860$ days for a Sun-like star), while all other works integrated over a radius $R_p$ range and single period $P$ or instellation flux $S$ range for all stars. We show the \cite{Bryson2021} estimates for two bounding completeness extrapolations. Estimates come from \citet{Petigura2013, ForemanMackey2014, Silburt2015, Burke2015}, SAG13, \citet{Mulders2018, Hsu2019, Pascucci2019, Bryson2020, Kunimoto2020, Bryson2021, bergsten2022}}
}
\label{fig:etaEarthHistory}
\end{center}
\end{figure*}

Many estimates of \etaearth have been calculated based on \kepler data.  Figure~\ref{fig:etaEarthHistory} shows a selection of \etaearth estimates to date and exhibits the large range of \etaearth values and uncertainties across these estimates.  These variations likely have several causes which we discuss briefly here, and are more fully examined in \cite{Bryson2025}.  These causes include:

\begin{itemize}
    \item Different studies adopt different \etaearth \textit{regimes} defined as ranges of planet size, orbital period or instellation flux.  These differing definitions are shown in Figure~\ref{fig:etaEarthHistory}.
    \item Studies make different choices of stellar parent populations, use different stellar properties, and use different exoplanet catalogs based on \kepler data.
    \item Studies use different algorithms and treatments of exoplanet catalog completeness (the fraction of true planets that are identified as planets in the catalog) and reliability (the fraction of planets in the catalog that are true planets).
    \item A very small number of earth-sized habitable-zone planet candidate detections due to their low S/N, leading to large uncertainties in \etaearth through small-number statistics.
    \item Extrapolation of population models to Earth-size planet populations from populations of larger planets.  The correct population model is unknown, and extrapolation of incorrect models leads to large errors.  Further, extrapolation is very sensitive to the details of the data being extrapolated.
    \item Because the Kepler project required three transits for a detection, the longest possible orbital period in \kepler catalogs is 710 days, while much of the HZ of stars hotter than about mid-G have orbital periods greater than 710 days. Thus extrapolation is required to cover all HZ of interest.  
\end{itemize}
These causes are discussed in more detail below.

As shown in Figure~\ref{fig:etaEarthHistory}, different studies adopt different definitions for the \etaearth parameter space, which can substantially change \etaearth estimates. For example, in \citet{Kunimoto2020}, simply lowering the minimum radius for what constitutes an ``Earth-size'' planet from $0.75$ to $0.5~R_{\oplus}$ can increase \etaearth by more than a factor of 2, because more planets are included in the \etaearth estimate from a lower-completeness region, resulting in larger completeness correction leading to more planets. While the results of \citet{Kunimoto2020} depend on the specifics of their population model and completeness correction, we believe that this large effect on \etaearth of changing the definition of ``Earth-size'' is typical. Different choices of HZ (typically the conservative or optimistic HZ) should also be reconciled before comparing values.  

Regarding differences in adopted catalogs, planet properties prior to 2018 were derived using stellar properties from the \kepler Input Catalog (KIC; \citealt{Brown2011}), with ground-based followup improvements on stars observed by \kepler \citep{Mathur2016}. After 2018, much improved stellar properties, particularly more accurate stellar radii, became available with data from \gaia combined with ground-based followup \citep{petigura2017california, Berger2020}. \citet{Bryson2020} compared using KIC and stellar properties from \citet{Berger2018} stellar properties for \etaearth defined as the occurrence of planets within 20\% of Earth's size and orbital period \cite[$\zeta_\oplus$ of ][]{Burke2015}. Using KIC stellar properties resulted in this \etaearth being 1.8 times higher, with 30\% larger error bars, than when using \citet{Berger2018} stellar properties. Different planet catalogs were used either because authors chose to create their own catalogs \citep{petigura2013prevalence,Kunimoto2020} or because they used differing \kepler catalog releases, with the list of planet candidates generally growing with later catalog releases. Studies from 2018 or prior used various \kepler catalog releases, while those after 2018 (with the exception of \citealt{Kunimoto2020}) used the final \kepler release DR25 \citep{thompson2018planetary}. Procedurally, when comparing \etaearth measurements, we recommend taking into account the definition of \etaearth and which stellar and planet catalogs were used.

While methods of computing \etaearth differ, one expects that ``good'' methods should agree on the value of \etaearth when using the same definitions and input data. A major difference between studies, however, is the treatment of the reliability of the exoplanet catalog, discussed below, which has only been carefully performed in studies after 2019.  \citet{Bryson2020b} studied four planet candidate catalogs created from the same input data and using the same method, differing only in vetting thresholds used to identify planet candidates.  They found that the SAG13 \etaearth computed from these four catalogs varied by a factor of 4.5 ($2 \sigma$) when failing to consider catalog reliability, but varied by a factor of 1.6 ($0.6 \sigma$) when accounting for catalog reliability. 

Differing definitions, catalogs and algorithms are primarily procedural issues and go a long way towards explaining the variations in \etaearth before 2019.  However, despite convergence on these issues, there is little evidence of convergence in estimates of \etaearth after 2019. We believe that this is due to the last two items in the above list: the lack of detections in the \etaearth regime, and the requirement to extrapolate due to incomplete coverage of HZ. These problems, discussed in more detail below, will be present in any future \etaearth study based on \kepler data or any data set whose detection limit is in the \etaearth regime.

\begin{figure*}[ht]
    \centering
    \includegraphics[width=0.95\textwidth]{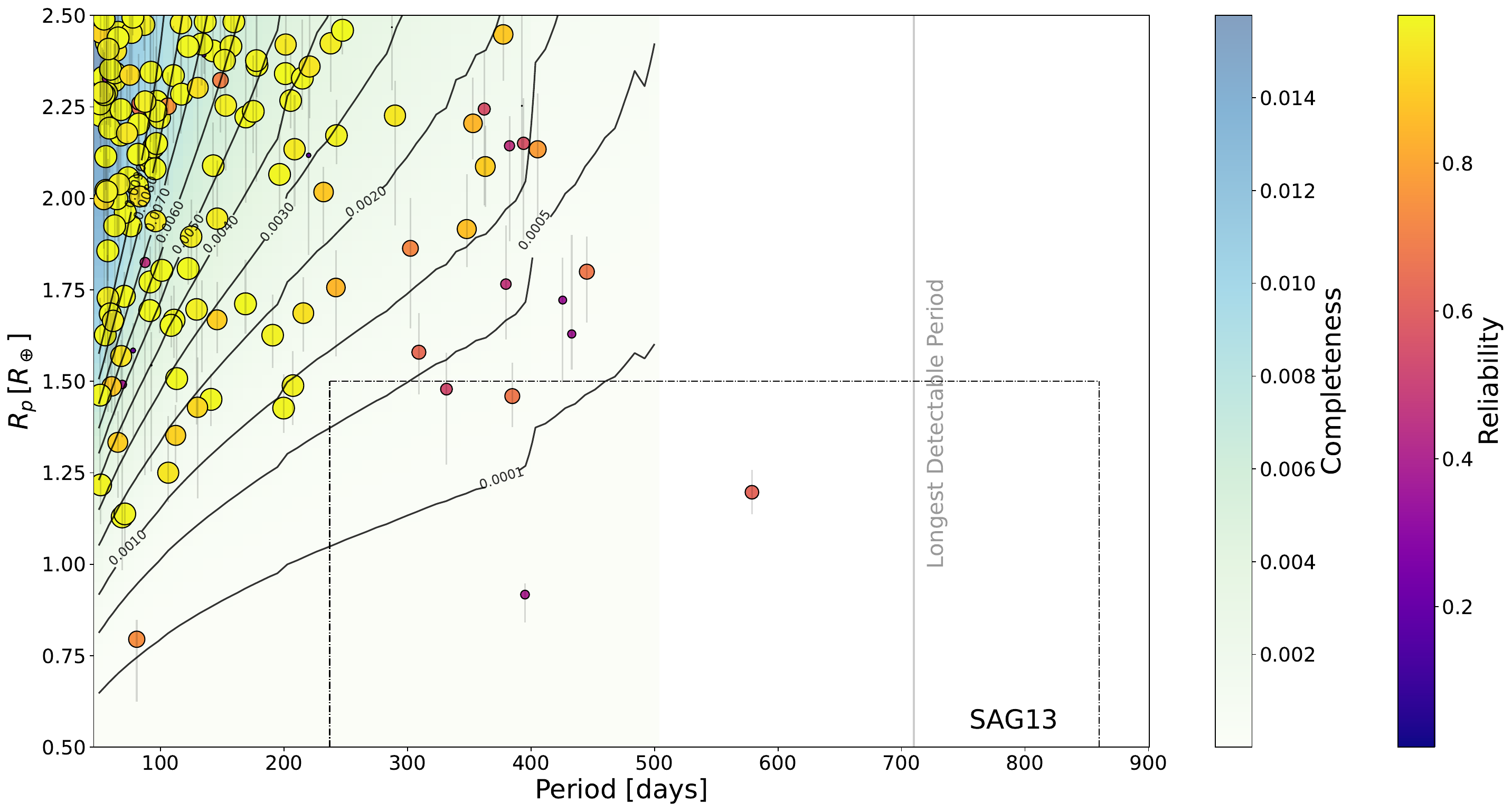} 
    \caption{ The DR25 Kepler exoplanet candidate population around FGK stars used in the analysis of \citet{Bryson2021}, shown in period and radius, colored and sized by reliability with exoplanet radius error bars.  The background color map and contours indicate detection completeness, which was only measured for orbital periods $<500$ days.  The longest detectable period from Kepler data assuming three transits is indicated at 710 days.  The rectangle shows the SAG13 definition of $\eta_\oplus$,  Adapted from~\cite{Bryson2020}. }\label{figure:populationPerRad}
\end{figure*}

While \kepler's photometric precision was very high, several factors resulted in very few detections of rocky planets in the habitable zone, and those few detections have low reliability \citep{thompson2018planetary}, for example see Figure~\ref{figure:populationPerRad}. The noise level of Sun-like dwarf stars was higher than expected by the \kepler team \citep{gilliland2011, Gilliland2015}, resulting in the need for an extended mission beyond 3.5 years to reach the original mission goals. However, the failure of two (of four) reaction wheels in 2012--2013 prevented such an extended mission from coming to fruition. Thermally-dependent electronic and other artifacts in the \kepler spacecraft also injected a very large number of signals that look very much like small rocky planet transits \citep{thompson2018planetary}.  Because thermal conditions repeat with the \kepler 372.5 day orbital period, these artifacts are easily mistakenly identified as small planets within the HZ of Sun-like stars.


For the final \kepler data release DR25, fully  automated detection \citep{Jenkins2010} and vetting \citep{Twicken2018, Coughlin2017} enabled characterization  of the completeness \citep{christiansen2016measuring,Christiansen2020} and reliability \citep{thompson2018planetary,Bryson2020} of the resulting planet candidate catalog.   However, the DR25 detection pipeline had very low completeness detecting small planets in the habitable zone of Sun-like stars.  On top of that, real planets may have been removed by the aggressive vetting needed to filter out the large number of artifacts in this regime.  While these planet losses can be accounted for via completeness measurements, adding or removing just one planet can have a dramatic effect on \etaearth, as will be demonstrated below.  The overall result is that the DR25 exoplanet catalog has very few detections of rocky planets in HZ of Sun-like stars, leading to a small-number statistics problem and resulting large uncertainties.  The lack of planet detections also leads to extrapolating from populations of larger radius or shorter orbital period planets (with more detections) to the \etaearth regime.


The need for two extrapolations, from larger to smaller planets and from shorter to longer orbital periods combined with few detections, is a significant source of uncertainty in measurements of \etaearth for at least two reasons:
\begin{itemize}
    \item We do not know what population function to extrapolate because we don't know the population demographic structure in the HZ of Sun-like stars. While (broken) power laws are a popular choice, there is significant evidence that exoplanet populations do not follow simple power laws, particularly for smaller planets \cite[e.g.,][] {fulton2017california,NeilRogers2020,Pascucci2019,bergsten2022}.
    \item Even if we knew the correct population model to extrapolate, extrapolations are strongly dependent on data at the boundary of the data being extrapolated, particularly when that boundary data is sparse. \citet{Bryson2025} discusses an example showing that adding or removing a single rocky habitable zone planet candidate can change \etaearth estimates by factors of 2 or 3.
\end{itemize}

To summarize this section, few planet detections and a lack of observational constraints on population models in the \etaearth regime make measuring \etaearth very challenging. Future observations, as well as deeper analysis of \kepler data, are expected to provide more detections which should improve our ability to measure \etaearth (see \citet{Bryson2025} for further discussion). Further constraints on \etaearth can be found via more sophisticated analysis discussed in the rest of this paper.

\section{Factors Not Fully Addressed in \etaearth Estimates}
\label{sec:missingAspects}
While Section~\ref{sec:etaEarthHistory} reviews key approaches to estimating \etaearth, several important factors that can significantly influence this measurement are not typically incorporated. In this section, we examine a set of these considerations and discuss their potential impact on our understanding of Earth-like planet occurrence rates.

\subsection{Stellar Multiplicity}
Previous estimates of \etaearth have focused on analyses of single stars, with attempts to completely remove binary stars from those samples. However, stellar multiplicity is a common outcome of star formation, with approximately 50\% of field FGK stars existing in binaries or higher-order multiples over all orbital separation down to mass ratios of 0.1 \citep{Duquennoy1991, Raghavan2010}. The multiplicity fraction is mass-dependent and increases with stellar mass \citep{duchene2013stellar, Offner2023}. Even the lowest-mass stars have multiplicity fractions around 25\% \citep[e.g.,][]{Kraus2012, Winters2019}. The presence of a binary companion alters the protoplanetary disk mass and lifetime \citep[e.g.,][]{Jensen1996, Cieza2009, Harris2012, Kraus2012}, suggesting that the planet population formed in binaries may differ from those in single star systems. The high frequency of binary stars makes them a major contaminant in many stellar samples and means that stellar multiplicity may substantially impact measurements of \etaearth. In addition, binary stars themselves may have different planet populations, and therefore different values for \etaearth.

Binary contamination in stellar samples is common because follow-up for stellar multiplicity to conclusively rule out a binary is time-consuming and resource-intensive. For large samples like the \kepler Input Catalog \citep[KIC;][]{Batalha2010} that are typically used for \etaearth measurements, it is not possible to rule out multiplicity of every source because intensive follow-up of every star is too expensive, meaning that binaries inevitably contaminate supposed single-star-only samples. Relevant to \etaearth, undetected binaries contribute flux that can dilute the transit depth of planet candidates, causing the planets to appear smaller than they truly are \citep[e.g.,][]{ciardi2015understanding, furlan2017kepler}. This in turn leads to over-estimations of a survey's sensitivity for detecting planets of any given size, which subsequently causes underestimations in the occurrence rate of that size of planet \citet{Savel2020}. Bergsten et al. (in prep) searched for unresolved stellar companions in a representative subsample of \kepler{} stars using adaptive optics, and implemented population-level corrections to gauge their effect on \etaearth. They found that correcting for unresolved companions and subsequent misestimations of survey completeness may lead to a 1.2--1.45 times relative increase in \etaearth, although such increases are often still consistent with original \etaearth measurements at $1\sigma$ given other sources of uncertainty.


The census of binaries in the \kepler sample of \kepler Objects of Interest (KOIs) is approaching completeness \citep[e.g.,][]{Kraus2016, furlan2017kepler, Ziegler2017,Sullivan2024}. In the \kepler target sample, which is most commonly used in calculations of \etaearth, the most complete assessment of potential stellar multiplicity at this point will come from a combination of the stellar characterization efforts like those of \citet{Berger2020}, who used HR diagram offsets to assess likely binaries, coupled with supplemental assessment of candidate binaries using \gaia DR3 \citep{Gaia2023DR3summary}. The sample of candidate binaries after selection in this way is approximately 10,000 candidate systems out of the total sample of $\sim$ 175,000 stars in \citet{Berger2020}, providing a lower limit on the total binary contamination fraction of $\sim$ 6\% (Sullivan et al. in prep.). 

Uncertain binary star contamination fractions are the major limiting factor to quantifying the impact of binaries on measurements of \etaearth. A complete binary star census does not exist for most samples. With \gaia DR3, many binaries are identifiable because they are angularly resolved, produce elevated astrometric noise (i.e., have an elevated Renormalized Unit Weight Error, RUWE above $\sim$ 1.4; \citealt{Lindegren2021}), or are spectroscopic binaries. However, when comparing the \gaia binaries to field star statistics like those of \citet{Raghavan2010}, the smallest-separation binaries remain incomplete even with the most comprehensive assemblies of candidate and known binaries in \gaia, and even the most reliable \gaia multiplicity metrics are not sensitive to all binaries because of limited astrometric precision and spatial resolution \citep[e.g.,][]{Sullivan2025gaia}. \gaia DR4, which will include astrometric solutions for binaries down to the spacecraft diffraction limit of 0\farcs15, as well as a longer time baseline for spectroscopic binaries, will be crucial for a more complete binary census in planet-host samples, although it will still not resolve all nearby binaries, because companions in face-on (non-spectroscopic) orbits can occur with very close orbits of fractions of au (or milliarcsecond separations).

\begin{figure}
    \centering
    \includegraphics[width=\linewidth]{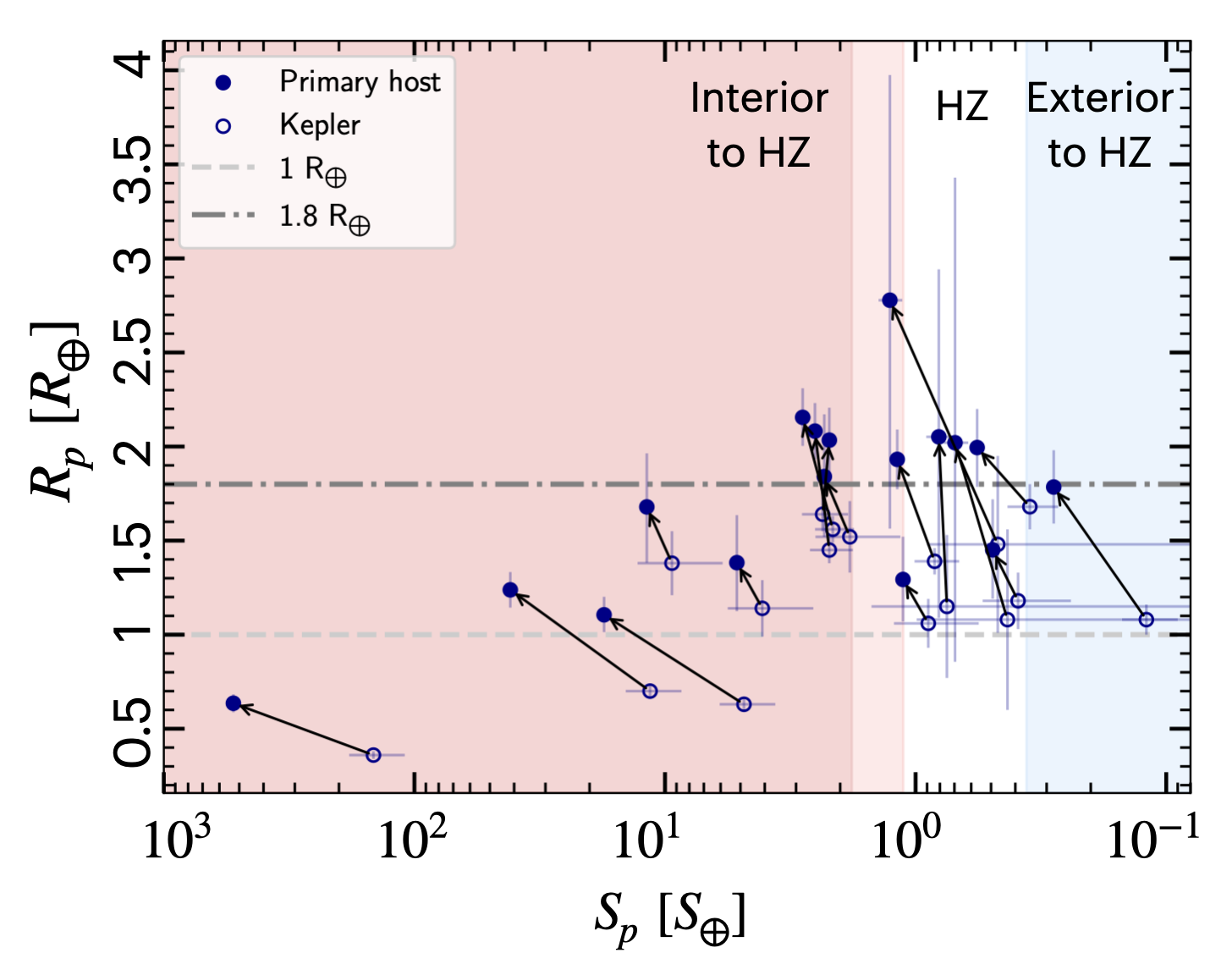}
    \caption{Planetary radius if the primary star is the planet host, plotted against the instellation flux for the planet. The open circles denote the \kepler DR25 values \citep{thompson2018planetary}, while the closed circles denote the revised values from \citet{Sullivan2022c} The dark red shaded region is the region inside the optimistic HZ, while the light red shaded region is the instellation range between the conservative and optimistic HZ (calculated using \citet{koskinen2013escape}). The blue shaded region falls outside both the conservative and optimistic HZs. The light and dark gray dotted lines show the 1 and 1.8 R$_{\oplus}$ boundaries, respectively. One of the planets originally outside the HZ moves into it. Almost all the planets in the HZ move above the radius gap, and 12 of the total 17 apparent super-Earths are actually sub-Neptunes, suggesting that the majority of detected planets in binaries are not true super-Earths. Figure adapted from \citet{Sullivan2022c}.}
    \label{fig:sullivan2022}
\end{figure}

In the \kepler Earth analog sample specifically, \citet{Sullivan2022c} estimated a similar binary contamination fraction for Earths and super-Earths of $\sim$5-10\%, based on their re-assessment of near-habitable-zone sub-Neptune interlopers that were misclassified as super-Earths before the system stellar multiplicity was discovered (Figure \ref{fig:sullivan2022}) Thus, binary contamination of \etaearth samples is likely at the $\sim$ 10-15\% level at minimum (from comparing known binaries to the California Kepler Survey sample of RV-observed stars; \citealt{Howard2010CPS}), which makes it a significant but not enormous source of contamination for single-star samples. Bergsten et al. (in prep) found an \etaearth\ binary contamination rate for the full \kepler\ sample of $\sim$ 20\%. Future \gaia data releases and additional high-resolution imaging follow-up for binaries in planet survey samples will be crucial for fully constraining the impact of binary stars on calculations of \etaearth. Adaptive optics (AO) and speckle imaging are especially important because they can reach separations and binary contrasts that \gaia is not sensitive to, but require dedicated follow-up time that is expensive for very large samples. These efforts must include work to constrain the parameter space where \gaia is sensitive to binaries to understand any remaining binary contamination.

In other samples using different observational techniques to constrain \etaearth, such as astrometric, radial velocity, or microlensing samples, the impact of binary contamination is nearly unconstrained (although see, e.g., \citealt{Hirsch2021} for an example of addressing multiplicity in demographics for the radial velocity method). Additionally, occurrence rates and \etaearth (among other demographics) remain difficult to measure in binary star systems, although work has suggested that the demographics of planets in binaries may indeed differ from those in single stars, potentially impacting \etaearth in binary star systems \citep[e.g.,][]{Wang2014, Kraus2016, Lester2021, Moe2021, Ziegler2021, Clark2022, Sullivan2023, Sullivan2024}.

Stellar (and sub-stellar) multiplicity has become easier to identify with \gaia and increasingly powerful high-resolution imaging capabilities, but binary stars remain a pernicious contaminant of single-star exoplanet demographic samples like those used to calculate \etaearth with transiting exoplanets. Not all binaries impact planet formation, but small- and intermediate-separation binaries, which include at least half of the binary star separation distribution \citep{Raghavan2010} have the largest impact on planet formation. \gaia DR4 will reveal more binaries, but high-contrast imaging and spectroscopic surveys will remain an important complement to identify as many binaries as possible in demographic samples. Studies of the occurrence rates and other demographics of planets in binaries remain in the early stages and are limited by small sample sizes, but results so far have shown intriguing hints of differences between populations of planets in single stars versus binaries.

\subsection{Planet Multiplicity}

Systems with multiple planets are also a common outcome of planet formation. While most of the systems with planets detected by \kepler (and even more so for those from the Transiting Exoplanet Survey Satellite (TESS) mission, \citealt{ricker2015tess,guerrero2021}; and other missions with shorter baselines) contain only a single transiting planet, roughly $\sim 40\%$ of all transiting planets observed by \kepler occur in multi-transiting systems (systems with more than one transiting planet; \citealt{Lissauer2011, Fabrycky2014, Lissauer2024}). The \kepler-observed planet-multiplicity distribution exhibits systems containing two, three, four, and even up to eight transiting planets, totaling several hundred of such multi-transiting systems \citep{Thompson2018, Lissauer2024}. When combined with the fact that transit probability and detection efficiency decrease with increasing orbital period, these observations imply that many single transiting systems likely also have unseen planetary companions, and detailed population modeling shows that these multi-planet systems are the norm (e.g., \citealt{FangMargot2012, Ballard2016, Mulders2018, Sandford2019, He2019, He2020}).

Even though multi-planet systems likely make up the vast majority of the underlying planetary systems, essentially all studies on \etaearth have not yet accounted for the role of planet multiplicity. While the Exoplanet Task Force Report \citep{Lunine2008} used the term ``fraction of stars" (with at least one potentially habitable planet) in their definition of \etaearth, the quantity that has been typically computed for \etaearth instead is a planet ``occurrence rate", which is defined as the mean number of planets per star. The distinction between an ``occurrence rate" (which quantifies an average rate, and thus can be greater than 1) and the ``fraction of stars with planets" (which by definition must be between 0 and 1) relies on accounting for planet multiplicity. While the former can be computed by treating planets independently (accounting for their individual detection efficiencies), computing the latter requires modeling the underlying multiplicity distribution.

Inferring the planet multiplicity distribution is especially challenging due to model degeneracies (as well as the added complexity of properly accounting for detection efficiency given multiple transiting planets; e.g. \citealt{Zink2019}). In particular, the number of planets in each system and the distribution of mutual inclinations between planets in the same system both strongly affect the number of transiting planets (e.g., \citealt{Lissauer2011, FangMargot2012, Johansen2012, Ballard2016, He2019, He2020}).
Furthermore, there is strong evidence that the distribution of mutual inclinations is anti-correlated with planet multiplicity \citep{Zhu2018, Yang2020}, which can be shown to arise from the angular momentum deficit (AMD) stability criteria (\citealt{He2020}; see also \citealt{Millholland2021}).\footnote{A similar anti-correlation also exists for the orbital eccentricity distribution \citep{He2020}. This is consistent with studies finding that planets in observed multi-planet systems tend to have lower eccentricities than those in single-planet systems (e.g., \citealt{zinzi2017, vanEylen2019}).} The relevant quantity for \etaearth would be the multiplicity distribution \textit{restricted} to the HZ. The width of the HZ generally increases towards hotter stars and is typically one to several AU for Sun-like stars \citep{Bryson2021}. Considering the separations relevant to the \etaearth regime, the ratio of the semi-major axes of the outer to inner HZ boundaries is a factor of several, large enough to potentially host multiple Earth-mass planets on stable orbits (for reference, the \kepler-observed period-ratio distribution for multi-planet systems peaks at $\lesssim 2$). Thus, estimates of \etaearth may potentially be affected by planet-multiplicity if it is often the case that more than one Earth-sized planet exists in the habitable zone of a single star.

Because the HZ of Sun-like stars are at/beyond the detection limits of \kepler, measuring \etaearth and multiplicity at these separations requires extrapolation, as emphasized in \S\ref{sec:etaEarthHistory}.
The extrapolation problem is further complicated by evidence for a truncation in the pattern of orderly, multi-planet systems that were found to be so common by the \kepler survey \citep{Millholland2022}, at the separations required for the \etaearth regime \citep[see also][]{NeilRogers2020}. \citet{Millholland2022} showed that if an additional planet (with a period and radius following the period ratio and planet size patterns of the inner planets) were to exist exterior to the outermost transiting planet in each of the \kepler 4+ systems, they would have been detected in $\gtrsim 35\%$ of the systems. Thus, the lack of such detections suggests an outer ``edge-of-the-multis" occurring at $\sim 100-300$ days, highlighting that one must be cautious when attempting to extrapolate towards longer periods from the compact inner multi-planet systems.
The combination of these factors makes it very tricky to account for planet multiplicity in estimating \etaearth even based on transit surveys, and substantial future work is needed to begin approaching this problem in order to gain an understanding of not only the frequency of Earth-sized habitable planets, but also how they are distributed across their host stars.


\subsection{Conditional occurrence rate of \texorpdfstring{$\mathcal{\eta_{\oplus}}$}{Eta-Earth}}

\begin{figure}
    \centering
    \includegraphics[width=\linewidth]{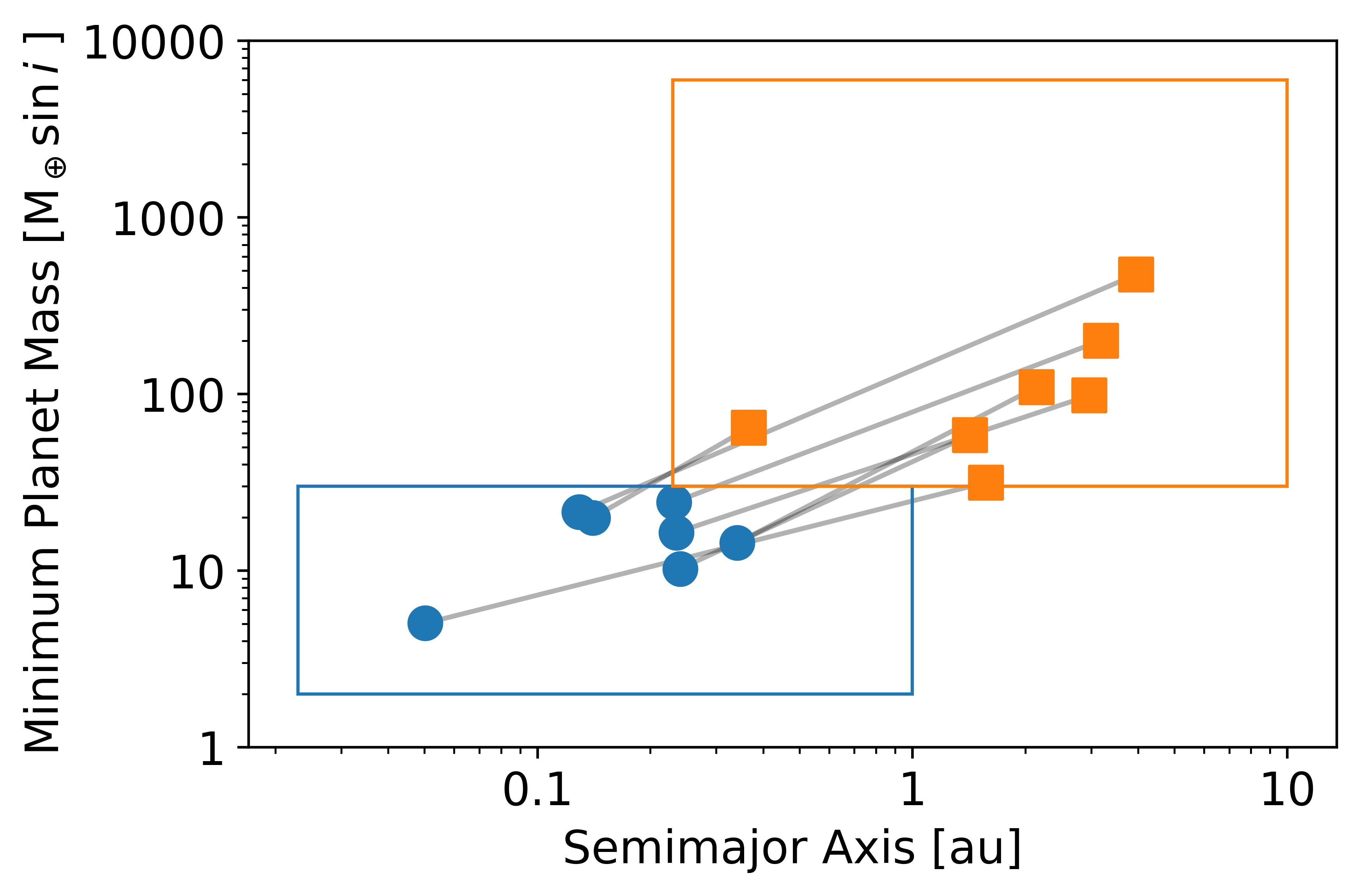}
    \caption{Studies of Earth-Jupiter correlations usually rely on a sample of small planets closer-in than the habitable zone. This figure shows the sample of systems (grey lines) with both inner super-Earths (blue circles) and outer giants (orange squares) from the California Legacy Survey \citep{Rosenthal2021,Rosenthal2022};  we exclude all but the first-detected planet of each type for each system. An additional example of known systems with inner small and outer giant planets can be found in Figure 4 of \citet{VanZandt2025}.}
    \label{fig:RosenthalSample}
\end{figure}

Conditional occurrence rates refer to a concept of occurrence rates within multi-planet systems, and provide a framework for asking what the occurrence of one type of planet is \textit{given} the presence of another type of planet in the system. Such questions are often motivated by the Solar System, where we seek to understand how common systems with Earths \textit{and} Jupiters are in an exoplanetary context. Due to a paucity of multi-planet systems with Earth-sized planets in their HZ detected in demographic surveys, there are no works on conditional occurrence rates that have provided \etaearth estimates.

Instead, studies to date exploring conditional pairs are usually looking at a generalized group of inner super-Earths and outer giants \citep[e.g.,][]{ZhuWu2018, Bryan2019, Rosenthal2022, Weiss2024, Bryan2024, VanZandt2025}, with varying definitions. For example, \citet{ZhuWu2018} define inner super-Earths as planets within 1 au and ``with mass/radius between Earth and Neptune,'' and outer giants as beyond 1\,au with $M > 0.3 M_J$; meanwhile, \citet{Rosenthal2022} define inner super-Earths as $0.023-1\,$au, $2-30\,M_\oplus$ planets and outer giants as planets with $0.23-10\,$au, $30-6000\,M_\oplus$. The planet sample from \citet{Rosenthal2022} is shown in Figure~\ref{fig:RosenthalSample} and demonstrates the lack of small planets beyond $\sim0.5$\,au in their sample of multi-planet systems, at least partially attributable to increasingly poor sensitivity for smaller and farther-out planets.

Because there are few (or no) systems with ``true" Earth- and Jupiter-analog pairs, a conditional \etaearth would require some combination of extrapolation and bounded integration. This in turn requires more complex/detailed descriptions of conditional occurrence beyond the scope or feasibility of previous works. Similarly, the presence of multiple planets in/near the habitable zone may affect the formation and habitability of inner small planets -- for example, which materials are available in the inner disk to form planets out of \citep[e.g.,][]{Lambrechts2019}, or planet-planet gravitational interactions influencing orbital stability. However, these relationships remain difficult to study observationally due to limited samples of both Earth- and Jupiter-analog planets; the latter is expected to improve with \gaia observations, discussed further in Section~\ref{astrometry}. In general, while there is growing interest in how \etaearth varies with planetary system architectures, a larger and more diverse sample of observed systems with Earth-like planets is likely necessary to better explore and characterize conditional occurrence rates affecting Earth-like planets.


\subsection{Chemical Composition}

Most studies constraining \etaearth have not considered the impact of host star abundance, which can affect the super-Earth occurrence rates \citep{Mulders2016Metallicity,petigura2018california,Zink2023,Boley2024}, formation \citep{Johnson2012,Lee2014,Lee2015,Lee2019}, architecture \citep{Bryan2019,Bryan2024, Zhu2024}, and interior and atmospheric properties that impact green house gases and thus the habitable zone and habitability \citep{Unterborn2019, Schulze2021,Adibekyan2021, Unterborn2023}. 

Host star abundances are thought to be a proxy for the available metals within the protoplanetary disk during formation \citep{Johnson2012,Hasegawa2014, Lee2019,Gilbert2025}. Given the relative ease of measurement compared to other abundances, metallicity has been used as a tracer for planet formation via core accretion \citep[e.g.][]{Johnson2012,Hasegawa2014, Lee2019, petigura2018california,Zink2023}. Early studies of the planet-metallicity correlation considered giant planets as they exhibit the strongest metallicity correlation of all planet types between  $-0.4 \leq$ [Fe/H] $\leq 0.4$ dex \citep{gon1997,FischerValenti2005, Udry2007, Johnson2012}. Small planets have a weaker correlation between their occurrence rate and host star metallicity compared to giant planets around solar and super-solar metallicities \citep[e.g.,][]{Buchhave2012,petigura2018california,Zink2023}. At sub-solar metallicities, super-Earth occurrence rates have a significant reduction below [Fe/H] $\sim$ $-0.4$ \citep{Boley2024}. At super-solar metallicities, super-Earth formation is positively correlated with the formation of an outer Jupiter \citep{Bryan2019,Bryan2024, Zhu2024}. In fact, super-Earth formation may be facilitated by a companion outer Jupiter at super-solar metallicities \citep{Bryan2019}. Most studies on the planet-metallicity correlation limit their samples to FGK stars. The samples are limited to FGKs due to survey designs, difficulty of detecting planets around OBA stars, and the challenge of measuring chemical abundances in M dwarfs. Therefore, more studies are necessary to determine the impact of metallicity on different spectral types. Given the impact of metallicity on super-Earth formation and system architecture, it is likely to affect calculations of \etaearth. 

While still in the preliminary stages, recent studies have begun to consider the effect of a variety of stellar abundances on super-Earth formation \cite[e.g.][]{Wilson2022,Rampalli2024}. Previously, exoplanet demographic studies were limited due to the lack of stellar abundance properties available. \cite{Wilson2022} studied the impact of ten different elements (C, Mg, Al, Si, S, K, Ca, Mn, Fe, and Ni) on the planet occurrence rates using a sample of $\sim 22,000$. They found a $\sim$20\% increase in the occurrence rate of super-Earths (1--1.9 R$_\oplus$) with orbital periods of 1--10 days around stars with an enhancement in any element of 0.1 dex. However, longer period super-Earths (10--100 days) showed a weaker trend with stellar abundance enrichment. Similarly, \cite{Rampalli2024} considered a sample of $\sim$ 17,000 stars and did not detect statistically significant trends between host star abundances and super-Earth formation within their precision limits. Better spectroscopic precision and larger samples are necessary to determine whether stellar abundances beyond metallicity are correlated with super-Earth formation at longer periods. However, galactic location and stellar age likely impact the role of stellar abundances on super-Earth formation, as they are correlated. A comprehensive study accounting for these factors would enable the field to disentangle stellar abundances from galactic properties and determine their impact on \etaearth.

To date, chemical compositions have not been considered in the calculation of \etaearth due to data availability of stellar abundances and precise planet masses ($\leq$ 30\% uncertainty). However, chemical compositions of rocky planets will likely impact \etaearth as it is commonly defined with respect to the boundaries of the HZ. Chemical compositions of rocky planets change the location of these boundaries as they affect the likelihood of a planet to have surface water. The presence and amount of surface water play a crucial role in conjunction with temperature in defining the habitable zone, as it dictates the operational greenhouse effect.   Rocky planets around host stars with super-solar metallicities form within protoplanetary disks that have reduced water due to the minerals formed \citep{Bitsch2016, Bitsch2020}. Given that surface water directly impacts the HZ of a planet \citep{Kodama2018,Kodama2019,Hayworth2020}, rocky planets around host stars with super-solar metallicities would have reduced HZ range by $\sim 0.2$ au for Sun-like stars \citep{Kodama2019}. Therefore, \etaearth calculations would be impacted.

Despite the critical role of stellar abundances on occurrence rate, architecture, and habitability, most \etaearth studies have not considered host star abundances. To better constrain \etaearth, it is critical to consider factors that likely impact the habitable zone. Given that the study of chemical abundances on the impact of rocky planet occurrence rates is in it's infancy, they likely act as another axis for \etaearth. Additionally, they are able to also provide an insight into the likely planet structure and ability to host surface water.



\subsection{Age}\label{sec:age}

The observed radius distribution of planetary systems reflects both evolutionary and dynamical processes: (A) migration, in which Neptune-sized planets form beyond the snow line and later move inward through interactions with the protoplanetary disk, remnant planetesimal belts, or other planets as well as through  dynamical interactions with distant stellar (or sub-stellar) companions and passing stars (thought to be rare), and tidal interactions between the planet and the host star. \citep{Bourrier2023,Venturini2020,Izidoro2022}, and (B) evolution driven by atmospheric mass loss mechanisms such as extreme ultraviolet (EUV) and X-ray (hereafter, XUV) photoevaporation and core-powered mass loss, which sculpt planetary radii over time. The timescales on which they operate serve to constrain the ages of planets with certain radii. Since disks only live for $\sim$10$\,\mathrm{Myr}$ (e.g. see \cite{ErcolanoPascucci2017} for a review, in-situ formation and gas disk migration would require these Neptunes to form and migrate within that timescale. Conversely, high-eccentricity tidal migration is thought to take up to a billion years ---ranging from just after disk dispersal to timescales longer than a typical stellar lifetime. 

With photoevaporation, XUV energy from the host star heats the upper atmospheres of planets, stripping away their atmospheres on a timescale of $\sim$100 Myr \citep{owen2013kepler, Pu_2015, Berger2020}, with continued mass loss at lower rates for longer. In the core-powered mass loss model, the energy for atmospheric loss is provided primarily by remnant thermal energy from a planet's accretion phase and occurs on a timescale of $\sim$1 Gyr \citep{Owen_2016, ginzburg2018core}, though the role of core-powered mass loss after an initial post-formation atmospheric "boil-off" phase (on the scale of $<1$ Myr) may be negligible \citep{Tang_2024}.
The two mechanisms may dominate in different regimes based on the location of the XUV penetration depth, related to the planet's radius \citep{owen_mapping_2023}. Since each mechanism occurs on its own timescale, if either photoevaporation or core-powered mass loss dominates for certain planet radii, this provides some constraining power on the ages of planets with those radii. However, both mechanisms may occur at different times for the same planet or even concurrently, introducing some degeneracy in the associated age constraint \citep{owen_mapping_2023}. 

\kepler's lower detectability towards smaller planets and large orbital periods resulted in the detection of just one Earth-size planet in the habitable zone of a solar analog \citep{borucki2018kepler}. As a result, the value of \etaearth can only be determined by extrapolations of the more abundant short-period, small ($<$1.8\,\Rearth) planets, which is likely increased by stripped cores produced by photoevaporation. To highlight the impact on \etaearth, a study by \citet{Pascucci2019} showed that \etaearth drops by a factor of $\sim$4$-$8 when excluding the population of short-period (within 10$-$30\,days) small  planets. As atmospheric mass loss processes are not efficient at large orbital separations \citep{owen2013kepler} e.g., in the habitable zone of solar analogs, it is important to understand the contamination of sub-Neptunes into the small short-period planet population in order to properly extrapolate that into the habitable zone and obtain a more reliable estimate of \etaearth. A similar conclusion was reached by \cite{NeilRogers2020} in a different way: by modeling Earth-size analogs as a combination of two populations (photoevaporated planets and ``born rocky'' planets), they showed that \etaearth extrapolations could have significant systematic uncertainties based on the idea that Earth-size planets detected by \kepler are dominated by potentially stripped planets. Since extrapolating to longer periods assumes that similar physical processes are at play, extrapolating the \kepler detections to longer periods past the physical processes that produce the ``radius valley'' (whatever the method of producing the radius valley) raises serious concerns \citep{Lee2022}. \cite{bergsten2022} showed that the abundance of these stripped planets comprised roughly ~half of Kepler's closest-in ($<$10 days) super-Earths, causing overestimates in extrapolations of Earth-sized planet abundance at habitable zone separations. They also introduced a framework to isolate the occurrence of (presumed) intrinsically rocky planets identified at longer periods ($>$50 days) where atmospheric loss mechanisms may be negligible.

In this context, it is also important to note that the HZ of a star evolves over time as stellar luminosity and temperature change. This has led to the introduction of the continuously habitable zone (CHZ) as a more nuanced metric. Two definitions are commonly used: a “sustained CHZ,” where planets remain in the HZ from the onset of habitability to the present day, and a “fixed-duration CHZ,” where planets are considered continuously habitable if they spend a minimum time (typically 2–4 Gyr) within the HZ, regardless of when they enter it \citep{Hart1979,Tuchow2020,Truitt2020,Tuchow2023}. These definitions introduce an additional layer of complexity when attempting to extrapolate occurrence rates from short-period planets to the HZ, particularly if habitability is assumed to require long-term stability within the zone.

\begin{figure*}[!htb]
    \centering
    \includegraphics[width=\linewidth]{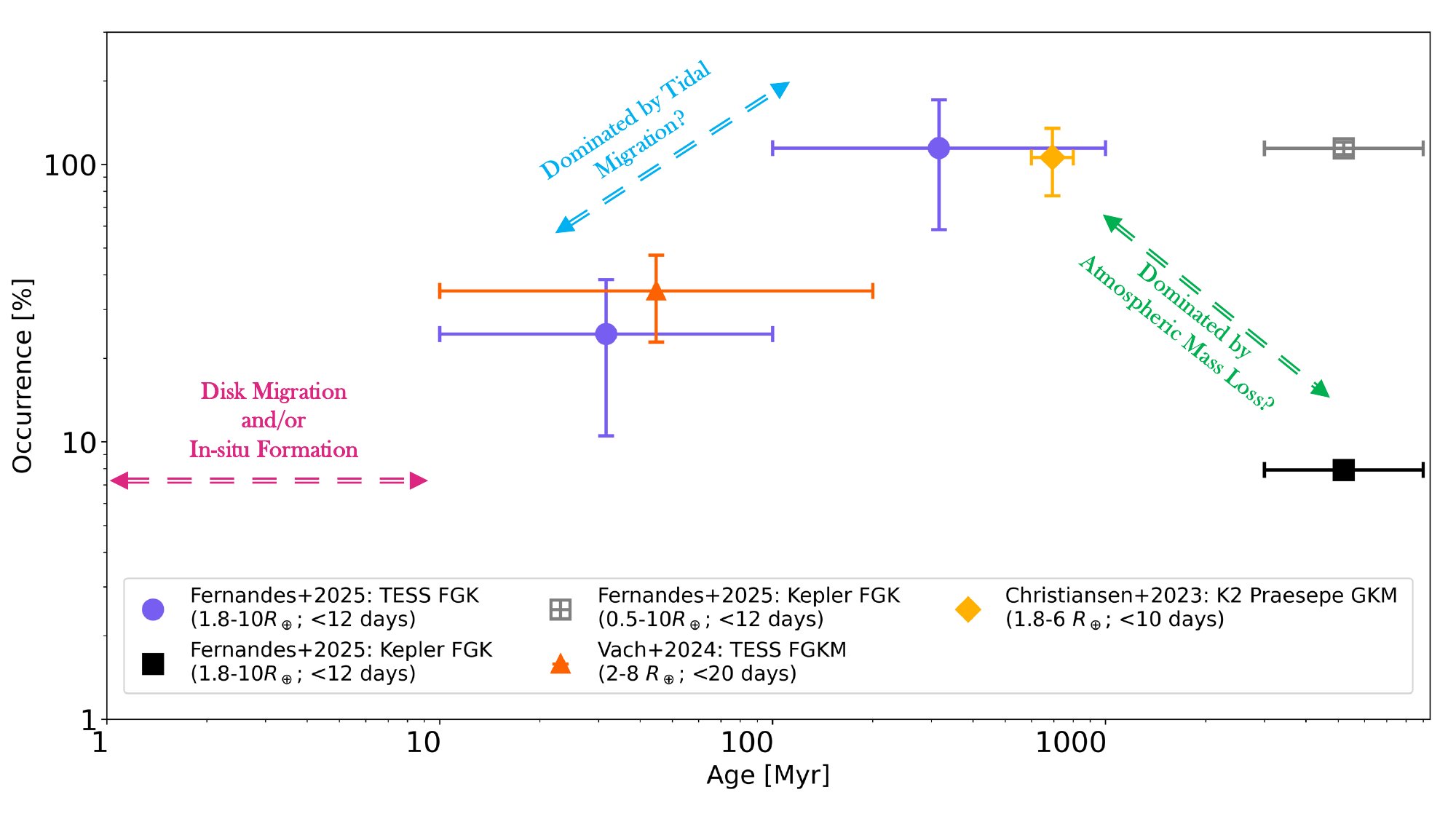}
    \caption{Comparison of occurrence rates for sub-Neptunes and Neptunes from different surveys exploring planet populations at different age bins, with the various underlying mechanisms and their timelines depicted with dashed arrows, from \cite{Fernandes2025_yc}}
\label{fig:occ_comp}
\end{figure*}

Understanding how planetary radii evolve over time is critical for interpreting the shape of the radius distribution at short periods—which, in turn, underpins the functional forms being extrapolated into the habitable zone to estimate \etaearth. Recent demographic studies on transiting sub-Neptunes in young clusters ($<$1 Gyr) with \ktwo \citep{Christiansen2023} and TESS \citep{fernandes2022,fernandes_khu_2023,Vach2024} have found that the occurrence of short-period sub-Neptunes and Neptunes is much higher than that of \kepler's mature population. Furthermore, \cite{Fernandes2025_yc} shows that the population of close in young planets is simultaneously being sculpted by both atmospheric mass loss mechanisms as well as inward migration (Figure~\ref{fig:occ_comp}). Currently, there are about 40 young planets found in both clusters/moving groups and young field stars, compared to thousands around Gyr-old field stars with \kepler, resulting in large uncertainties driven by the small sample size of young planets. Increasing the sample of young planets will enable a more statistically robust comparison of the intrinsic occurrence rates with time. There is a need to simultaneously compute the occurrence of longer-period, young planets ($>$12\,days), and compare it with that of short-period young planets to fully understand the role of post-disk migration. To evaluate if atmospheric mass loss is a dominant driver in the radius evolution of gas-rich planets we need to detect planets smaller than 1.8\,\Rearth (super-Earths) and compare their occurrence with that of gas-rich planets over time. Given that the light curves of young stars are highly variable, making it challenging to detect small planets, we need more precise light curves, which upcoming missions like PLATO \citep{Rauer2024} can likely provide.


\subsection{Galactic Population} \label{subsec:galacticpopulation}

\begin{figure*}[!htpb]
    \centering
    \includegraphics[width=1\linewidth]{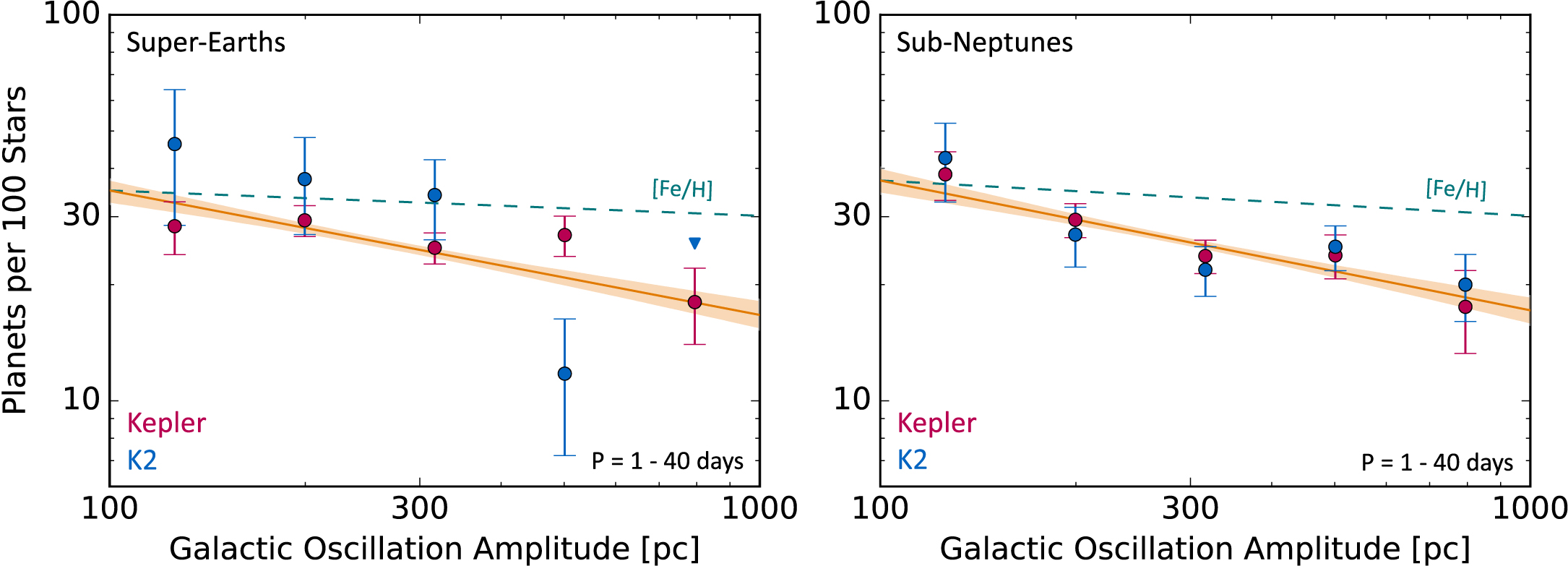}
    \caption{Planet occurrence vs. galactic oscillation amplitude for super-Earths and sub-Neptunes from \citet{Zink2023}. The red and blue points represent the raw planet occurrence within bins of galactic oscillation amplitude for Kepler and K2 planet hosts, respectively. The teal dotted line represents the expected planet occurrence rate from a calculated stellar metallicity gradient with galactic oscillation amplitude. While the relationship between planet occurrence is not yet well-constrained for planets of any size, this figure serves as a preliminary example of how \etaearth may scale with the stellar host's location and galactic orbit.}
    \label{fig:zink_galactic_occurrence}
\end{figure*}

Efforts to constrain \etaearth within a Galactic context are in their preliminary stages, since \gaia has made astrometric data for planet-hosting stars available. Planet occurrence and properties are strongly sensitive to stellar metallicity, and potentially to local stellar number density as well: for example, \citet{Gilliland_2000} found that planet occurrence in the globular cluster 47 Tucanae is potentially much lower than in the \textit{Kepler} field as a whole, though these results have been disputed \citep{Masuda_2017}. More recent studies have investigated the relationship between stellar clustering in position and velocity space and planet demographics. Though it is difficult to disentangle the effect of stellar clustering with related properties (e.g. stellar age and metallicity), these results demonstrate a potential planet—Galactic population trend and the potential for the combination of \textit{Gaia} and large-scale exoplanet survey data to investigate this idea. 

Theoretical studies suggest that stellar clustering may influence planet occurrence rates \citep{Shara2016,Wang2020, Benkendorff2024}. Several observational studies have worked to constrain the impact of stellar clustering on planet occurrence rate for Jupiters, finding little correlations \cite[e.g.,][]{Adibekyan2021A&A,Blaylock-Squibbs2023,Kontiainen2025}.  However, \citet{Longmore2021} found planet multiplicity to vary with stellar clustering, where stars in phase space overdensities are more likely to host single-planet systems and hot Jupiters. This could suggest that Earth-sized planets are more likely found in lower than average phase space density areas of the Galaxy, though disentangling Galactic kinematic properties with the intrinsic properties of stellar hosts (such as the stellar age) remains a difficult problem. \citet{mustill_alexander_j_hot_2022} find that the Sun itself lies in an area of higher than average stellar phase space density, though the lack of extrasolar analogs for Solar System planets makes it difficult to use this as supporting evidence for \etaearth as a function of phase space density.

\citet{Dai2021} found that main-sequence stars with high relative velocities have a lower occurrence rate of super-Earths and sub-Neptunes ($1-4$ \Rearth) and a higher occurrence rate of sub-Earths ($0.5-1$ \Rearth). This suggests that environmental perturbation linked to stellar clustering may significantly affect planetary orbits and their habitability. In the same trend, \citet{Zink2023} and \citet{Chen2021} found that the occurrence rate of super-Earths and sub-Neptunes is anti-correlated with the vertical Galactic velocities of their host stars (Figure \ref{fig:zink_galactic_occurrence}), while \citet{bashi_exoplanets_2022} find such a correlation with Galactic disk association. All this suggests that Earth-sized exoplanet hosts may be preferentially found in certain Galactic regions. However, more work is needed to precisely constrain \etaearth in the context of the Galaxy, especially as a function of Galactic height and vertical velocity. Furthermore, careful analysis is needed to disentangle Galactic properties from correlated stellar properties such as age and metallicity for main-sequence, Earth-hosting stars.


\subsection{Spectral Type}\label{sec:spty}

\begin{figure*}
    \centering
    \includegraphics[width=1\linewidth]{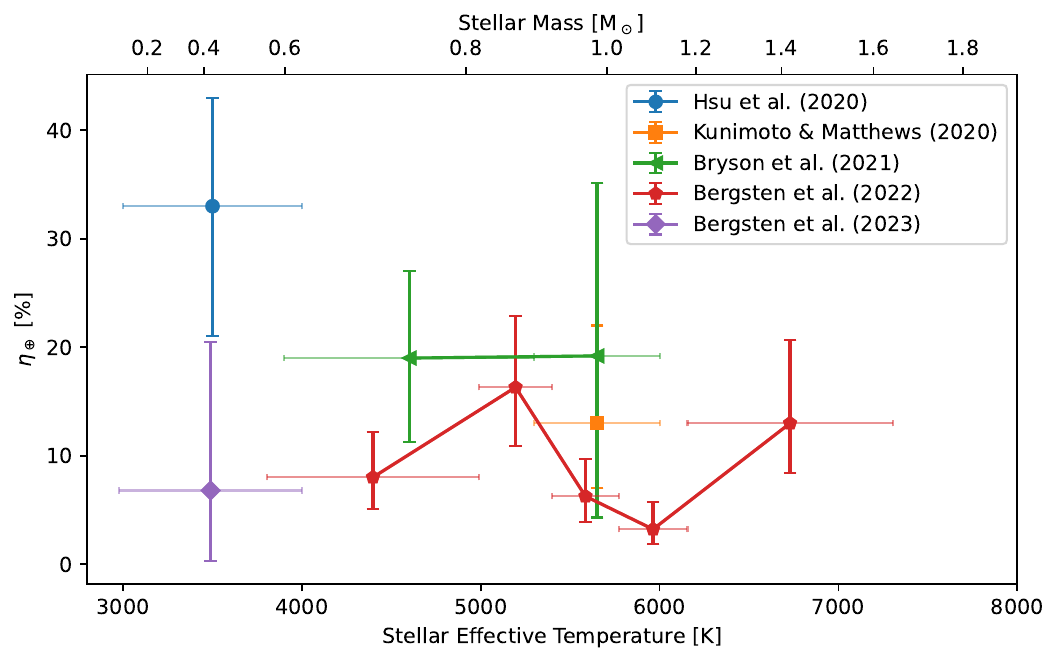}
    \caption{Literature measurements of $\eta_\oplus$ for various stellar effective temperatures and masses, highlighting the current lack of clear trends between/across spectral types. Vertical error bars represent uncertainties on $\eta_\oplus$, while horizontal error bars denote the range of stellar effective temperatures/masses considered for each study. Values shown use the same definition of $\eta_\oplus$ (0.75--1.5\,R$\oplus$ and the conservative habitable zone from \citealp{Kopparapu2013}), as reported in \citet{Hsu2020, Bergsten2023} for M dwarfs and \citet{Kunimoto2020,Bryson2021,bergsten2022} for Sun-like stars. We exclude works that may use different \etaearth definitions in order to provide the most direct comparison; for example, we do not include $\eta_\oplus$ estimates from \citet{dressing2015occurrence} which use 1.0--1.5\,R$\oplus$ planet radius bounds. While \citet{Bryson2021, bergsten2022, Bergsten2023} originally also used different planet radius ranges (0.5--1.5\,R$\oplus$, 0.7--1.5\,R$\oplus$ and 0.5--1.5\,R$\oplus$, respectively), we re-evaluated their occurrence models to provide $\eta_\oplus$ estimates over the 0.75--1.5\,R$\oplus$ range for this figure. Additionally, we note that this figure includes the conservative habitable zone estimate from \citet{Bryson2021}, while Figure~\ref{fig:etaEarthHistory} includes the optimistic one.}
    \label{fig:eta_spty}
\end{figure*}

The occurrence of small close-in (P$<100$\,days) planets tends to increase with decreasing stellar mass from A to K stars \citep[e.g.,][]{mulders2015stellar, Yang2020, He2021, bergsten2022, Giacalone2025}. This trend also increases into the M dwarf regime, where small close-in planets may be $\sim3.5$ times more common than around FGK stars \citep{mulders2015planet, Petigura2022, Hsu2020}. However, it is unclear whether these trends for short-period planets might translate to the longer-period habitable zone. Due to limited samples, \etaearth estimates for FGK stars are usually computed in a single stellar mass bin; as such, there are few measurements of how \etaearth changes from F to K stars.

Any attempts to model \etaearth in different bins of stellar effective temperature or mass are often hindered by large uncertainties or overlapping upper limits \citep[e.g.,][]{Hsu2019}. Figure~\ref{fig:eta_spty} shows a sample of \etaearth measurements (all using the same definition of 0.75--1.5\,R$_\oplus$ planets in the conservative habitable zone) across the FGKM regime, with no clear trends in how \etaearth should vary by spectral type. The occurrence models of \citet{Bryson2021} incorporated a dependence on stellar effective temperature and recovered a slight occurrence increase towards cooler stars. Using this, they calculated separate \etaearth estimates around K and G stars (3900-5300\,K and 5300-6000\,K, respectively), but the corresponding uncertainties were too large to draw significant conclusions (e.g., $\eta_\oplus = 0.49_{-0.26}^{+0.51}$ versus $0.60_{-0.34}^{+0.75}$ for an optimistic habitable zone). Similarly, \citet{bergsten2022} did not find any clear trends in \etaearth across FGK stars (0.56-1.63\,\Msun) despite recovering the rise in small close-in planet occurrence with decreasing stellar mass. This is attributable to variations and uncertainties in their other model parameters, namely the occurrence-period slope for planets with orbital periods larger than $\sim15$\,days.

Uncertainties in the properties of the host stars themselves must also be considered. \citet{tayar2022} finds fundamental error floors of $\sim2\%$ in stellar effective temperature and $\sim4\%$ in stellar radii due to current levels of uncertainty in bolometric fluxes and interferometric measurements, to which stellar evolution/isochrone models are often calibrated. They also find systematic uncertainties between stellar model grids of $\sim5\%$ in stellar mass and $\sim20\%$ in age for main-sequence stars. These are important effects for stars of all spectral types when determining the planetary demographic properties of inhomogeneously measured host stars.

Furthermore, it is somewhat common for the $\sim$3--4$\times$ increase in close-in planet occurrence from FGK to M dwarfs to be applied to \etaearth, despite the former only being measured for planets with short orbital periods \citep[e.g.,][]{HardegreeUllman2023}. We note that \etaearth is conventionally focused on Sun-like stars, but the same definitions from Section~\ref{sec:intro} can be applied to M dwarfs for a non-interchangeable parameter (often still labeled as \etaearth) specific to these lower-mass stars. \kepler{}'s already sparse measurements of small planets at long orbital periods, combined with the mission having observed so few M dwarfs, makes it difficult to discern how \etaearth scales with stellar mass between M and FGK stars. \citet{Bergsten2023} found that the \kepler sample lacked sufficient evidence to support measurements of any \etaearth trends with stellar mass across FGKM stars, calling for more observations from other surveys to improve M dwarf constraints. Generally, because measurements of \etaearth are already sample-limited, attempts to further divide observed samples by spectral (sub-)type tend to suffer from larger uncertainties. This obscures any potential trends in how \etaearth should vary by spectral type (Fig.~\ref{fig:eta_spty}), so more detections of Earth-like planets across \textit{all} spectral types would promote a more clear understanding.

\subsubsection{Examining \texorpdfstring{$\mathcal{\eta_{\oplus}}$}{Eta-Earth} for M dwarfs}

\begin{figure*}
    \centering
    \includegraphics[width=\linewidth]{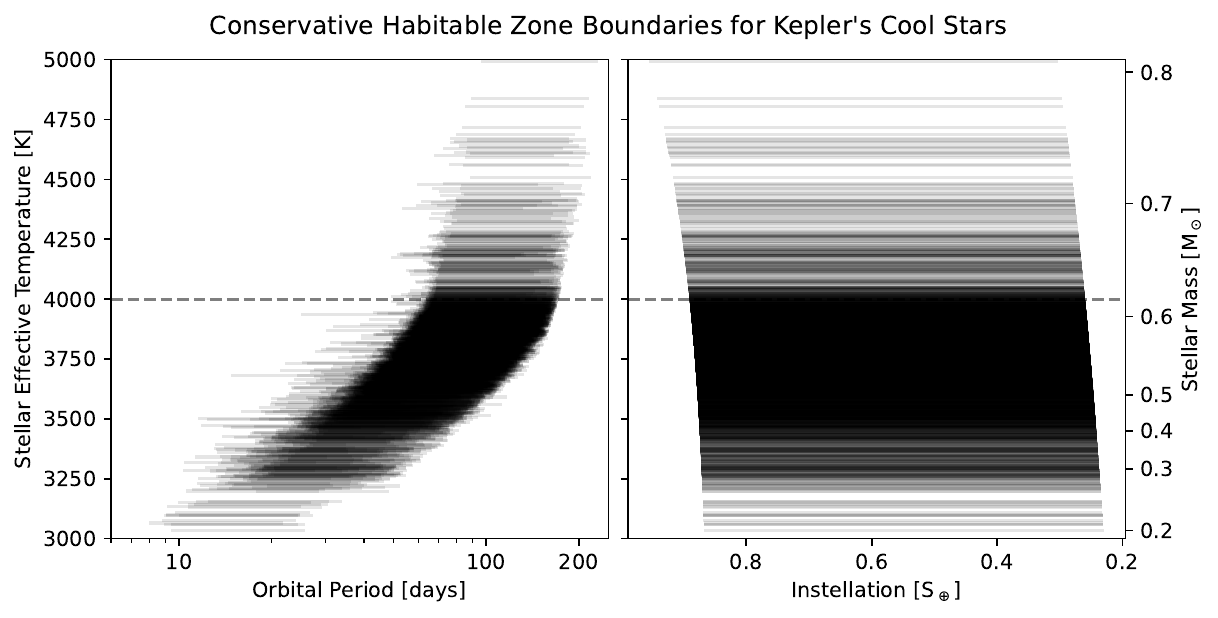}
    \caption{M dwarfs span a wide range of stellar effective temperatures, so their HZ vary strongly in orbital period, but are roughly constant in instellation. This plot shows the conservative habitable zone boundaries for \kepler{}'s cool stars as a function of stellar effective temperature (or stellar mass). Solid horizontal lines denote an individual star's habitable zone (from \citealp{Kopparapu2013}) plotted in \textbf{left:} orbital period or \textbf{right:} instellation. The dashed line at 4000\,K marks a typical upper boundary for the M dwarf spectral type \citep[see e.g.,][]{dressing2015occurrence}. The instellation boundaries being roughly constant for M dwarfs makes it easier to integrate habitable zone occurrence rates, compared to the temperature-dependence of orbital period boundaries necessitating either an average habitable zone or (more accurately) a temperature-dependent integration.}
    \label{fig:HZ_bounds}
\end{figure*}

In terms of finding Earth-like planets, M dwarfs have some advantages over Sun-like stars that could benefit a refined understanding of \etaearth -- which, in this section, will refer to Earth-sized planets in the habitable zones of M dwarfs -- specific to these lower-mass stars. Their smaller sizes and cooler temperatures can increase the feasibility of detecting Earth-sized, habitable zone planets, and their status as the most abundant spectral type in the solar neighborhood makes them a popular choice when searching for nearby planets. Compared to FGK stars, there is also an extremely wide range of stellar properties that constitute the ``M dwarf" spectral type --- roughly $0.6$\,\Msun down to $0.1$\,\Msun, or a range of about $1500$\,K in effective temperature. Such a broad spectral type means the location and width of the habitable zone can vary wildly in orbital period (e.g., widths ranging anywhere from $10$ to $100$ days, \citealp{Hsu2020, Bergsten2023}; see Figure~\ref{fig:HZ_bounds}). Thus, while most FGK studies evaluate occurrence rates in orbital period, it is more common for M dwarf studies to operate in instellation space where the habitable zone boundaries have much less variation (see e.g., \citealp{Kopparapu2013}). 

There are likely several confounding variables that may affect planet habitability around M dwarfs more so than around Sun-like stars -- e.g., orbital eccentricity, tidal heating, flares, etc. In particular, we highlight the issue of atmospheric mass loss (as discussed in Section~\ref{sec:age}) and the fact that M dwarfs tend to remain XUV-active for longer than their FGK counterparts. In turn, this can exacerbate issues of contamination when measuring small planet occurrence rates (see e.g., the increased ratio of super-Earths to sub-Neptunes around M dwarfs, \citealt{CloutierMenou2020,MentCharbonneau2023}), and/or suggest that ``Earth-sized" planets in the HZ of M dwarfs could be more likely to be stripped of their atmospheres and thus less likely to be habitable. However, accounting for these issues when studying M dwarf planet occurrence rates is difficult in the already information-limited studies of \etaearth. 

\citet{dressing2015occurrence} was a seminal study of \kepler's M dwarfs, reporting $\eta_\oplus = 16^{+17}_{-7}\%$ for a conservative habitable zone and $24^{+18}_{-8}\%$ for an optimistic habitable zone. However, recent advancements -- most notably \gaia stellar property revisions \citep{Berger2020} -- now labeled many stars from the original \citet{dressing2015occurrence} sample as K dwarfs, with planets too large or hot to fall within standard \etaearth boundaries (as noted in \citealp{Bergsten2023}). \citet{Hsu2020} incorporated \gaia parameter revisions in an occurrence rate forward model to derive $\eta_\oplus = 33^{+10}_{-12}\%$ for $[0.75,1.5]\,R_\oplus$ planets in the conservative habitable zone. \citet{Bergsten2023} also incorporated \gaia revisions and found a value of $\eta_\oplus = 9^{+18}_{-8}\%$, consistent at $1\sigma$; their large uncertainties are driven by the paucity of reliable detections of habitable zone Earth-sized planets in the current \kepler sample.

While the TESS time baseline for an individual sector is typically too short to explore M dwarf HZ, stacking data from multiple sectors that observed the same patch of sky (through either repeat observations or the Continuous Viewing Zone) potentially offers a long enough multi-sector baseline to probe this topic. As such, longer period studies may be possible with TESS as the community's demographic capabilities continue to develop -- but so far, most TESS studies rely on single sectors and can thus only probe out to $\sim$10 days. The work of \citet{MentCharbonneau2023} suggests a sensitivity of roughly $50\%$ to $1 R_\oplus$ planets at 7 days; this corresponds to roughly 4 times Earth's instellation for their sample of mid-to-late M dwarfs. Nevertheless,  this promotes hope of some non-zero sensitivity at the longer orbital periods necessary to search the habitable zone. This likely does not apply to the latest M dwarfs observed with TESS, where increased photometric noise in TESS observations of less massive (thus generally fainter) M dwarfs may cause the mission to miss tens of rocky planet candidates \citep{Brady2022}.

For transits, precise properties for the M-dwarf host stars themselves are essential in determining the bulk properties of their planets. However, radii of M-dwarfs are especially difficult to estimate, and are consistently determined from eclipsing binary and photometric/SED observations to be $\approx$5\% larger than the radii predicted by stellar models \citep{Chabrier2000,Torres2014,Morrell2019}. This discrepancy has been called the ``radius inflation problem," and provides an acute reminder that occurrence estimates around M dwarfs are particularly sensitive to methods of stellar characterization. The Eclipsing Binary - Low Mass (EBLM) Survey is increasing the sample of measured M-dwarf masses, radii, and effective temperatures through characterizing single line eclipsing M-dwarfs with an FGK type host through transit and radial velocity observations \citep{Triaud2013, Swayne2024}. Using the CARMENES radial velocity instrument at the Calar Alto Observatory \citep{CARMENES2014}, \citet{Schweitzer2019} have measured the fundamental properties of several hundred M-dwarfs through high resolution spectroscopy. As the sample of precisely characterized M-dwarfs increases, empirical relationships to estimate key M-dwarf properties like mass and radius \citep{Mann2015,Mann2019,Pineda2021} will continue to improve.


Radial velocity surveys of M dwarfs have also struggled with sparse detections of habitable zone candidates. Recently, \citet{Pinamonti2022} placed an upper limit of $\eta_\oplus < 23\%$. This is consistent with the estimate of $21^{+3}_{-5}\%$ for 3--10\,\Mearth~from \citet{Tuomi2014}, and a re-evaluated estimate of $\sim20\%$ from \citet{Bonfils2013} after updates to exclude a false positive (see Section 6.4 of \citealp{Pinamonti2022} for more details). Regardless of the method, more detections of Earth-like planets around M dwarfs are likely necessary to better constrain \etaearth around these abundant low-mass stars, and further clarify how \etaearth may change between FGK and M dwarf stars as discussed in Section~\ref{sec:spty}.


\section{Detection Techniques and The Path to \etaearth}
\label{sec:futureProspects}
While most detection techniques are, in principle, capable of finding Earth-like planets, each comes with inherent limitations that affect their sensitivity to such planets in practice. In this section, we examine how these limitations influence our ability to constrain \etaearth, evaluate the regions of parameter space where current techniques are most and least effective, and consider how upcoming missions plan to address these challenges.

\subsection{Transits}


As discussed in section \ref{sec:etaEarthHistory}, the best estimates of \etaearth derive primarily from the \kepler transit survey, which has yielded several disparate estimates due to a combination of procedural issues and the paucity of planet candidates detected in the habitable zone (see also Fig.~\ref{fig:etaEarthHistory}). 
The paucity of candidates can ultimately be traced back to hardware failures that limited the lifetime of the \kepler mission to a relatively short duration ($\sim$1400 days) relative to that needed for a secure detection of a habitable zone planet transiting a Sun-like star with periods ranging from $P \approx 240$-860~days, depending on the adopted habitable zone definition.  

Future transit surveys can improve upon the strides made by the \kepler mission in a number of ways. In particular, by meeting three separate criteria (see also \cite{winn2024} for a review on occurrence rates in transit surveys). 
First, a survey must monitor stars over a long enough baseline, and with few enough gaps in observing, to observe at least 2-3 transit events. 
For Sun-like stars, the required baseline to guarantee this criterion is met equates to $\gtrsim$3 years, though three transit events could be detected in as little as a 2 year baseline. 

Second, a hypothetical survey must monitor stars at a sufficient photometric precision 
to detect transit events from $\sim 1 R_\oplus$ objects. For Sun-like stars, this transit depth and duration is $\sim$80 ppm and $\sim$8-12 hr, respectively. 
Thus, to detect an Earth-sized planet with 3 transit events at $S/N = 10$, the photometric precision must be better than 15~ppm when averaged over the approximately 8-12 hr duration of the transit, or better than 45 ppm$\,\rm{hr}^{1/2}$ assuming white noise. 
However, this limit begins to approach the level of intrinsic stellar variability for FGK stars, which was found to range from $\sim$10-30 ppm over transit-like timescales in \kepler \citep{gilliland2011}. This noise manifests as stochastic short-period ($<$8 hr) brightness variations in the light curves of F, G, and K stars \citep{bastien2013,bastien2016}. The dominant variations are likely caused by granulation, the actively evolving patterns of cool and warm spots on the surfaces of convective stars \citep{cranmer2014,vankooten2021}. In addition to granulation noise, pressure-mode oscillations are also present at $\gtrsim$10 ppm amplitudes and with a periodicity on timescales of several minutes for Sun-like stars. 
Together, these noise sources present challenges for \etaearth estimates for Sun-like stars because they can inhibit the detection of shallow transits and significantly bias planet radii measurements inferred from transit models \citep{sulis2020}. Thus, future transits surveys aiming to refine \etaearth for FGK stars will need to account for these effects 
by incorporating stellar variability in their transit modeling and 
by budgeting for mission lifetimes longer than the minimum $\sim$2-3 yr baseline needed to observe enough individual transit events in order to mitigate stellar variability noise floors.

\begin{figure*}
    \includegraphics[width=\linewidth]{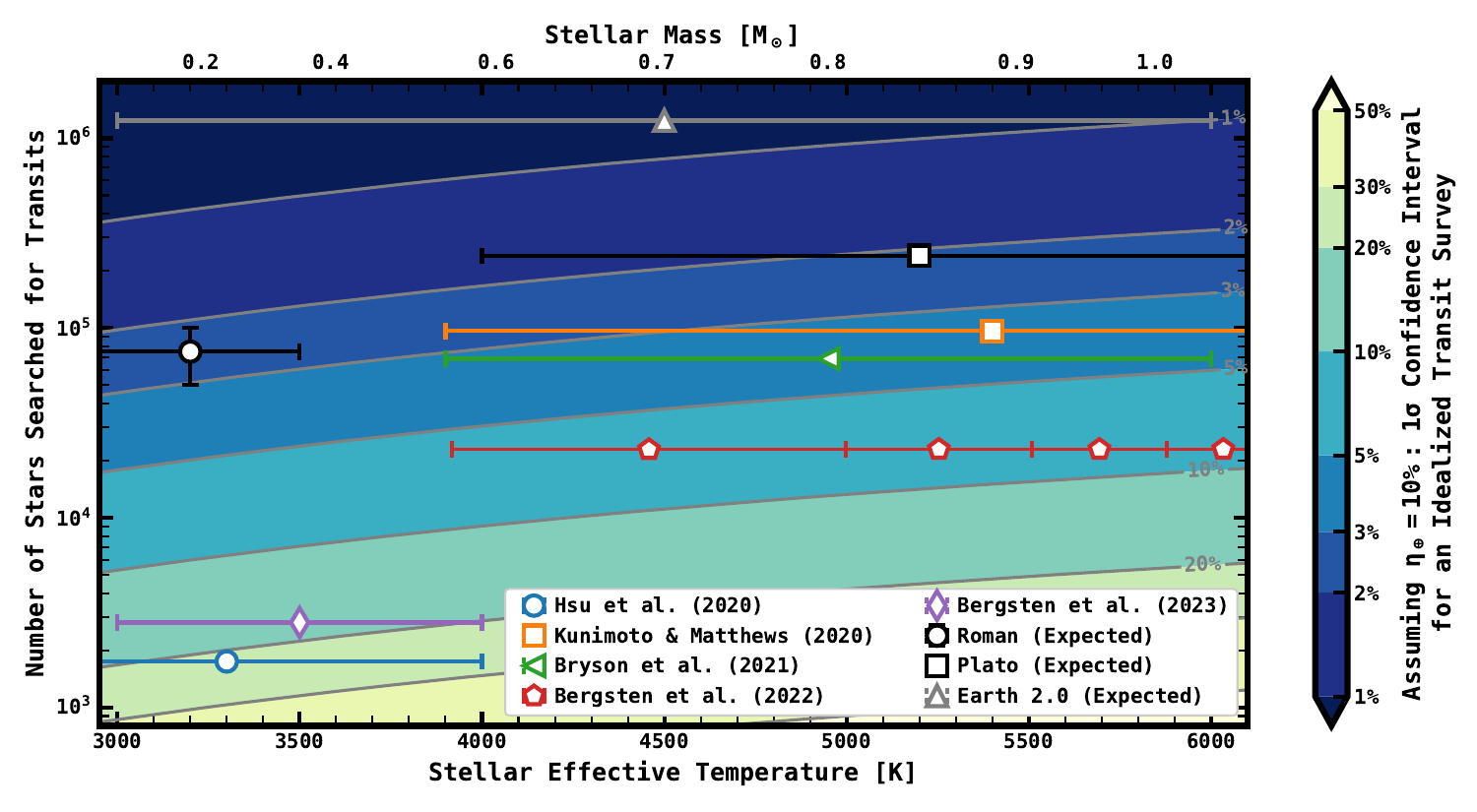}
    \caption{The minimum possible uncertainty in \etaearth, represented as the 68\% confidence interval, for a transit survey set by the geometric transit bias, $p_{\rm tra}$, as a function of stellar type and number of stars searched. Plotted over the contours are \etaearth studies from \kepler showing their sample size and range of spectral types considered, as well as expected sample sizes from upcoming transit missions. The habitable zone bounds are calculated from \cite{Kopparapu2013} and the stellar properties are drawn assuming a 4 Gyr isochrone \citep{choi2016,dotter2016}, a typical age for an FGK star in the \kepler field \citep{Berger2020}.}
        \label{fig:eta_earth_confidence}
\end{figure*}

Lastly, a transit survey must monitor a sufficient number of stars to overcome the geometric transit bias, $p_{\rm tra}$. This can be expressed as $p_{\rm tra} =  \frac{R_\star}{a (1-e^2)} \approx \frac{R_\star}{a}$. For planets orbiting Sun-like stars at a distance of 1 au, $p_{\rm tra} \approx 0.5\%$. To predict the uncertainty on \etaearth as a function of stellar type and number of stars searched for transiting planets ($n_\star$), we model an idealized transit survey as a binomial distribution, where the number of effective trials corresponds to $p_{\rm tra}\times n_\star$ and $p_{\rm tra}$ is determined by marginalizing over the inner and outer  bounds of the habitable zone from \cite{Kopparapu2013}. With a conservative  assumption of \etaearth=10\%, we calculate the minimum uncertainty due to $p_{\rm tra}$ by calculating the Clopper-Pearson integral \citep{newcombe1998}, shown in Figure \ref{fig:eta_earth_confidence} across a range of stellar spectral types. 
Assuming \etaearth=10\%, a transit survey targeting Sun-like stars observed under perfect survey conditions (i.e., the rate of false positives, false alarms and false negatives are all equal to 0), would need to observe $\gtrsim$330,000 stars to infer \etaearth with a 68\% confidence interval to better than a factor of 0.1 (i.e., \etaearth$\approx~10\%^{+1\%}_{-1\%}$), or $\gtrsim$10,000 stars to infer \etaearth to within a factor of two (i.e., \etaearth$\approx~10\%^{+10\%}_{-5\%}$). 

However, it is unlikely that a transit survey could be achievable at such a high efficiency due to a number of factors that would increase the false positive and false negative rates, as described in previous sections (e.g., stellar variability).
As seen in Figure \ref{fig:eta_earth_confidence}, this is the case for \kepler studies of FGK stars, where the limiting factors are due to a combination of photometric precision and observing baseline, rather than sample size.
Therefore, in practice, the above sample size estimates typically serve as a significant lower bound on the number of stars needed to better constrain \etaearth via future transit surveys.

For lower mass stars, 
all three of these requirements can be relaxed for a number of reasons. First, stellar variability noise floors due to granulation and pressure-mode oscillations in low-mass stars are significantly lower than the expected transits depths of Earth-sized planets. Second, the HZ are at smaller semi-major axes and therefore planets have a larger $p_{\rm tra}$ and shorter orbital periods (see Fig. \ref{fig:HZ_bounds}), making detection significantly easier in theory. E.g., for a M5V star ($M_\star\approx0.2 M_\odot$), the minimum sample sizes would be $\gtrsim$110,000 and $\gtrsim$3000 to infer \etaearth within a factor of 0.1 and two, respectively, and the baseline need only be a few months to observe a significant number of transit events. In fact, as seen in Fig. \ref{fig:eta_earth_confidence}, the minimum uncertainty on \etaearth is comparable to the derived uncertainty on \etaearth for M dwarfs \citep{Hsu2020,Bergsten2023}, implying that for low mass stars, \kepler's estimate of \etaearth is largely limited by sample size. 
For this reason, a great deal of ongoing work to improve \etaearth is focused on lower mass stars.

At the time of this manuscript, much of the ongoing effort to constrain \etaearth via transit comes from the TESS mission. Due to TESS's small aperture (10 cm), \etaearth constraints will be limited to relatively nearby, luminous M dwarfs which have less stringent photometric requirements. 
The TESS observing strategy also provides a challenge. Because the TESS observing strategy is non-uniform, some parts of the sky have significantly more data than others, leading to large variations in the completeness and reliability of the TESS transit search from star to star. 
Thus, transit detection biases must be evaluated empirically for any arbitrary stellar sample. For this reason, occurrence rate studies can be challenging, particularly for areas of low detection efficiency. Despite these challenges, a few habitable-zone transiting planets have already been discovered and characterized with TESS data, including TOI-700e/d \citep{gilbert2020, rodriguez2020, gilbert2023} and LP 890-9c \citep{delrez2022}. The latter transiting planet was discovered initially from the ground-based SPECULOOS survey \citep{sebastien2021}, which suggests that ground-based transit surveys may be a promising strategy for constraining \etaearth for ultracool dwarf stars (M7 and later). 

The earliest upcoming mission with the potential to constrain \etaearth for mid to late M dwarfs ($M_\star \approx 0.08$-$0.30\;M_\odot$) via transit is likely NASA's next flagship mission, the Nancy Grace Roman Space Telescope (Roman), set to launch in late 2026. 
As one of its Core Community Surveys, Roman will conduct a high-cadence (12.1 min), near-infrared ($\lambda = 0.9$-$2.0\;\mu$m) survey of a 1.68 square degree region near the Galactic Bulge, known as the Galactic Bulge Time Domain Survey (GBTDS). 
Due to the near-infrared bandpass and deep magnitude limits
($H\lesssim 21\;\rm{mag_{AB}}$), simulations of transiting planet yields indicate that 
the GBTDS is expected to monitor $\sim$50,000-100,000 M3-9V stars with the photometric precision needed to facilitate the detection of $\sim$10-40 transiting planets with $R_p \lesssim 1.5 R_\oplus$
in the conservative habitable zone defined in \cite{Kopparapu2013}. Applying these same simulations to the optimistic habitable zone in \cite{Kopparapu2013}, the expected yield may be as large as $\sim$120 \citep{tamburo2023,wilson2023}. These predictions are based on the occurrence rates calculated in \cite{dressing2015occurrence}. 




In addition to Roman and the efforts from TESS, there are upcoming missions with the goal of improving \etaearth for Sun-like stars specifically. For example, the PLanetary Transits and Oscillations of stars mission \citep[PLATO;][]{Rauer2014,2024arXiv240605447R} from the European Space Agency aims to detect transiting terrestrial planets out to the habitable zone for Solar-type host stars with an estimated launch date of late 2026. PLATO will observe approximately 245,000 FGK dwarf and subgiant stars across two fields, one in the southern hemisphere and one in the northern hemisphere, with observations taking place over at least 4 years. PLATO will focus on bright stars ($<$11 mag in the PLATO bandpass which ranges from $\lambda$=0.5-1.0~$\mu$m), which are well suited to radial velocity follow up observations. PLATO will also exploit high cadence observations to study high-priority host stars using asteroseismology to provide detailed characterizations of these systems. PLATO will include a 25 second high cadence observation mode for this purpose. Adopting lower and upper bounds for \etaearth from \cite{Bryson2021}, \cite{Heller2022} estimates that PLATO will discover between 10 and 30 systems with 0.5 \Rearth$<R_{p}<$ 1.5 \Rearth\ in the habitable zone around FGK stars.

Yet another upcoming mission expected to improve our current estimates of $\eta_\oplus$ is Earth 2.0 \citep[ET;][]{Ge2022,zhang2022}, expected to collect data in the 2030s. The ET mission has the stated objective of discovering Earth-like ($R_p$=0.8-1.25 $R_\oplus$) in the HZ of solar-type stars. The ET science payload consists of seven 30 cm telescopes, six of which will conduct a transit survey and one of which that will conduct a microlensing survey in the Galactic bulge, observing a total of approximately 1.2 million FGKM stars. 
ET is expected to discover 10-20 Earth-sized planets in the habitable zone over its 4-year mission lifetime.


\subsection{Radial Velocity}
\begin{figure}
\includegraphics[width=0.45\textwidth]{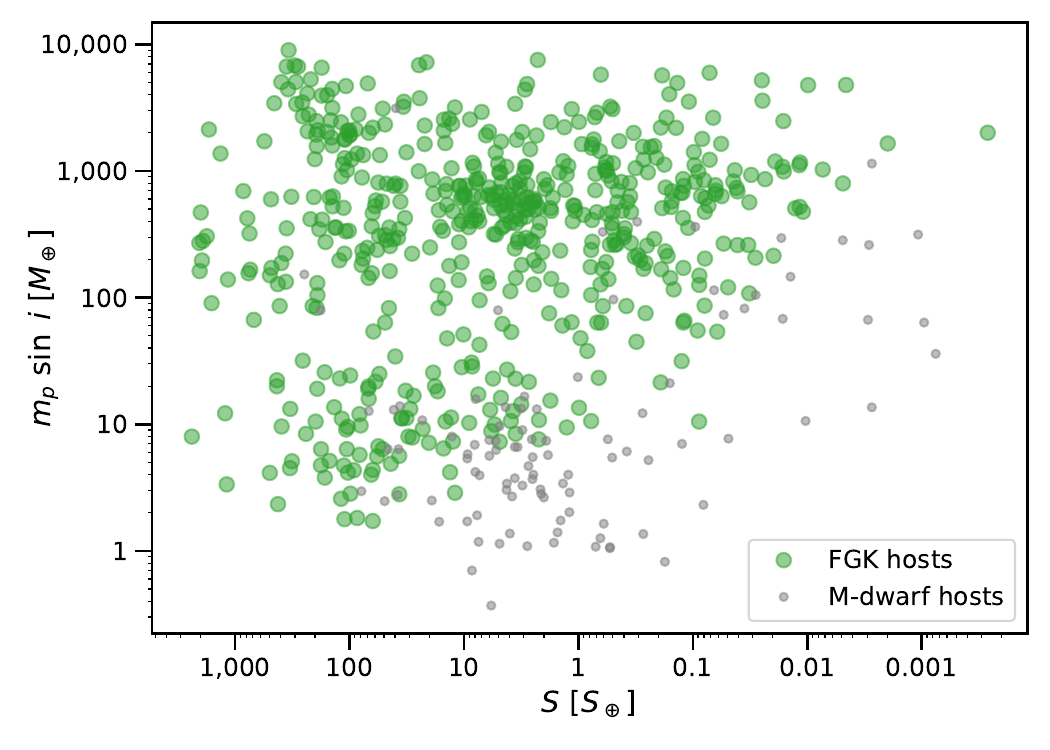} 
\caption{Minimum planet mass vs instellation for all planets discovered via the RV method. This sample is insufficient to constrain \etaearth for FGK host stars. Data taken from the NASA Exoplanet Archive \citep{Christiansen2025b}.}
\label{fig:rv}
\end{figure}

The RV method (see e.g., \citet{Lovis2010} for a review) is responsible for the discovery of more than 1,000 exoplanets, second only to the transit method. Constraints on planet occurrence rates from RV observations predate even the first exoplanet discoveries, with upper limits established in the Jovian mass regime for small samples of stars monitored by early RV surveys \citep{Campbell1988,Cochran1994,Walker1995}. 
As these and other surveys progressed, and as RV exoplanet detections ramped up, more robust occurrence rates could be measured. Spectrographs with internal precisions of 1--3 m~s$^{-1}$ enabled the analysis of trends with planet properties such as mass and orbital period \citep[e.g.,][]{Cumming2008,Howard2010,mayor2011,Wittenmyer2011}. But these studies were limited in scope by the achievable RV precision; the RV signal induced by an orbiting exoplanet is directly proportional to the planet mass and scales inversely with orbital period, and a true Earth analog ($1 M_\oplus$, 365 d, orbiting a 1 ${\rm M}_\odot$ star), induces a maximum signal with a semi-amplitude of just 10 cm~s$^{-1}$. \citet{Howard2010} extrapolate down from the super-Earth mass range ($3-10\ M_\oplus$) to derive an occurrence rate of 23\% for Earth-mass ($0.2-2\ M_\oplus$) planets with orbital periods of $<50$ days, but a measurement of \etaearth at HZ separations remained well out of reach for this generation of instruments. Even with an additional decade of data, sensitivity limits from these spectrographs do not encroach on the \etaearth mass-period regime \citep{Laliotis2023}. As we show in Figure~\ref{fig:rv}, no Earth-mass planets with Earth-equivalent instellation have been detected around FGK host stars with the RV method. In addition, unlike transiting exoplanet searches, RV surveys historically have not been carried out in a manner conducive to robust modeling of the search parameters, which in turn degrades the accuracy of any demographics results.

Prospects for measuring \etaearth with RV observations hinge on the success of the latest generation of surveys, including the HARPS-N Rocky Planet Search \citep{Motalebi2015}, the ESPRESSO blind survey \citep{Hojjatpanah2019}, the EXPRES 100 Earths survey \citep{Brewer2020}, and the NEID Earth Twin Survey \citep[NETS;][]{Gupta2021}, all of which are being carried out on extreme precision radial velocity (EPRV) instruments with demonstrated sub-m~s$^{-1}$ internal precisions. A recent report from the EPRV working group \citep{Crass2021,Luhn2023} notes that intrinsic stellar variability at the m~s$^{-1}$ level will still present a major obstacle to the detection of 10 cm~s$^{-1}$ planet signals. Still, RV observations have the distinct advantage that the facilities and infrastructure needed to measure \etaearth are already in place; some studies are optimistic that, with an improved understanding of stellar variability, ambitious and carefully designed surveys can enable occurrence rate measurements in the Earth analog regime \citep[e.g.,][]{Gupta2024}. But the presence of additional planets in Earth-hosting systems may introduce additional complications, as it becomes much more challenging to characterize the observed RV time series when multiple Keplerians must be disentangled
\citep{He2021b,Murphy2025}. We also note that occurrence rate estimates from RV-only planet detections are limited by the mass-inclination degeneracy. An Earth-mass planet on an edge-on orbit ($i=90^\circ$) will induce the same signal as a higher mass planet on a less inclined orbit. This degeneracy can be accounted for in a statistical manner with a large sample size, but it can not be broken for individual systems without a measurement of the orbital inclination using a different observing technique, such as astrometry.

\subsection{Imaging}

Imaging searches for exoplanets have revealed several young super-Jupiters on wide orbits \citep{Currie2022}. While  imaging methods progress toward detecting lower-mass planets, older planets, and planets on shorter orbital periods, constraints on the frequency of rocky planets (and thus on \etaearth) will be made possible. With current telescopes, rocky planets can be imaged around $\alpha$ Centauri in very long ($\sim$100 hrs) exposures \citep{Kasper2019,Wagner2021,Hunziker2020}. While rocky planets could presently be detected orbiting $\alpha$ Centauri, they are not ruled out, and Earth-sized planets are not currently detectable in attainable exposures.
With the upcoming 30-m class telescopes, it will be possible to image rocky Earth-sized planets within $\alpha$ Cen in a matter of hours \citep{Brandl2021,Kasper2021}. Rocky planets around several other nearby stars could be imaged in longer exposures \citep{Bowens2021,Werber2023}. In other words, constraints on \etaearth will become possible with ground-based imaging, but will still be severely limited by small sample statistics. For example, in the event that Earth-sized planets are ruled out around the seven closest FGK stars ($\alpha$ Cen A \& B, 61 Cyg A \& B, $\epsilon$ Eri, $\tau$ Ceti, Procyon), \etaearth will be constrained to $\leq$35\% (95\% confidence level, CL). Should one Earth-like planet be detected, \etaearth will be constrained to $14^{+38}_{-13}$\% (95\% CL). Of course, such initial constraints would not be more precise than those from Kepler. Also, stellar multiplicity will also need to be taken into account  (e.g., \citealt{Moe2021}). 

In the slightly more distant future, imaging will play a crucial role in determining \etaearth with with the Roman Space Telescope, high contrast for future ELTs, and purpose-built flagship-class space telescopes designed to search for Earth-like planets and biosignatures. The Habitable Worlds Observatory (HWO), as recommended by the Astro2020 Decadal Survey, has as a goal that the observatory ``will provide a robust sample of $\sim$25 atmospheric spectra of potentially habitable exoplanets, will be a transformative observatory for astrophysics, and given optimal budget profiles it could launch by the first half of the 2040 decade.'' Astro2020 adopted \etaearth = 0.24 (\cite{Bryson2020}; see Fig. 7.6 caption of Decadal), where they defined the HZ "as 0.95-1.67 au for planets of 0.8-1.4 Earth radii". Hence, this would suggest that HWO should be built to survey approximately $\sim$100 nearby stars for exoEarths. \cite{Mamajek2024} provided a preliminary catalog of the nearest target stars whose hypothetical exoEarths, would be not only on the most accessible orbits in terms of angular separation, but be the brightest. They posted a list of 164 stars in three tiers, with $\sim$100 of the stars considered the best available targets known at the time, in order to seed precursor science research to inform potential science cases and inform science yield simulations to inform HWO's design.
 
In the extreme case that no Earth-like planets are detected within this sample, such a survey would place an upper limit (95\% CL) on \etaearth $\leq$2\%. If ten Earth-like planets are found, then \etaearth would be constrained to $6^{+4}_{-2}$\% (95\% CL). HWO will operate in the visible to near-infrared with a high-performance coronagraph. Ground-based thermal infrared imaging (for very nearby systems: \citealt{Bowens2021,Werber2023}), or a dedicated space mission such as the Large Interferometer for Exoplanets \citep{Quanz2022}, would enable more detailed characterization (in particular, fully-constrained temperature, radii, and albedos) of these planets. 

\subsection{Microlensing} 
A microlensing event is a purely gravitational phenomenon that occurs when two astrophysical bodies become nearly perfectly aligned. In this configuration, the gravitational field of the foreground (lens) object bends the light of the background (source) star, resulting in an apparent transient magnification of the background star \citep{Gaudi2012}. This magnification does not rely on any emission by the lens. Thus, microlensing can be used to detect astrophysical bodies with electromagnetic emissions well below detectable thresholds, providing the potential to observe planets around stars fainter and farther than other methods allow.

The magnification caused by a single-lens microlensing event \citep{Paczynski1986} can be significantly perturbed if there is another massive object in the lens system. It is through these perturbations that microlensing surveys are able to detect planets orbiting a lensing star. In contrast to the transit and RV methods, microlensing is primarily sensitive to planets with intermediate orbital periods; maximal sensitivity occurs when the planetary perturbation falls near the effective lensing radius (``Einstein ring radius'') of a lensing star, which is of order 2-4 au for events with sources in the Galactic Center. As a result, microlensing studies of exoplanet occurrence rates uniquely enable the study of low-mass, wide-separation planets. Even beyond binary-lens systems (i.e., one planet and one star), which are the primary avenue for detecting planets via microlensing, the technique is sensitive to a broad range of system architectures, such as planets hosted by stellar binaries \citep[e.g.,][]{Bennett2016, Han2024}, multi-planet systems \citep[e.g.,][]{Gaudi2008, Han2013, Gould2014}, and planets unbound to any star \citep[e.g.,][]{Mroz2017}.

However, due to the precise alignment needed for microlensing to produce an observable magnification, microlensing events are exceedingly rare. To maximize the probability of observing such a precise alignment, microlensing surveys target areas of the sky with high stellar density, with the Galactic Center being the prime target area for detecting microlensing events in the Milky Way. Decades-long ground-based microlensing surveys targeting the Galactic Center, such as the Optical Gravitational Lensing Experiment \citep[OGLE,][]{Udalski1992}, Korean Microlensing Telescope Network \citep[KMTnet,][]{Kim2016}, and Microlensing Observations in Astrophysics \citep[MOA,][]{Muraki1999}, have provided some of the best measurements of the occurrence rates of wide-orbit planets to date \citep[e.g.,][]{Suzuki2016,Poleski2021,Zang2025}.
Ongoing observations by these collaborations 
are continuing to improve our understanding of such worlds.

The fundamental trade-off in microlensing is between observational area and cadence. Typically, microlensing surveys image multiple fields to maximize the number of sources, returning to each one at a fixed interval. Adding more fields increases the number of sources, but it limits how quickly the telescope can return to any given field. However, the duration of an event is proportional to $\sqrt{M_\text{lens}}$, with typical timescales on the order of hours for Earth-mass lenses. Hence, in order to detect the perturbations induced by such lenses, high-cadence observations are required. Only recently have ground-based surveys achieved the cadence necessary to begin to be sensitive to Earth-mass planets, however this sensitivity is still marginal. Detection is additionally complicated by intrinsic degeneracies in microlensing observables like the event duration, which is degenerate in the lens mass, lens distance, and lens proper motion. In most cases, microlensing light curves can only be used to infer star-planet mass ratios and projected separations of planets scaled to the Einstein ring radius. Physical properties of the planet can be inferred from microlensing parameters with additional observables such as lens flux \citep[e.g.,][]{Bhattacharya2018}, or higher-order light curve effects such as microlensing parallax \citep{Hardy1995} and finite source effects \citep{Witt1994}, if observed.

The launch of the Nancy Grace Roman Space Telescope, NASA's next flagship mission, will usher in a new era for microlensing. As the first dedicated space-based microlensing survey, it will observe the Galactic Bulge at sub-fifteen minute cadence and extraordinary photometric precision, enabling the detection of roughly 1400 bound planets through microlensing \citep{Penny2019}. One of its science requirements is, through these observations, to ``estimate \etaearth (defined as the frequency of planets orbiting FGK stars with mass ratio and estimated projected semimajor axis within 20\% of the Earth-Sun system) to a precision of 0.2 dex via extrapolation from larger and longer-period planets'' (EML 2.0.5 of Roman Space Telescope Science Requirements). While current estimates of \etaearth are made from the extrapolation of shorter period planets in transit surveys, microlensing can perform a similar extrapolation from planets at larger separations, providing a complementary means to estimate \etaearth. Furthermore, as stated above, the sensitivity to detecting a planet in the habitable zone via microlensing peaks when the habitable zone overlaps with the Einstein radius of the primary lens. In the context of Roman and other surveys targeting the Galactic Center, this occurs for host FGK stars for lens systems situated roughly 7.5 kpc from Earth. As a result, microlensing surveys provide a unique opportunity to constrain \etaearth for stars near the Galactic Center and thus a distinct population from the Solar neighborhood. If the differences in these two populations can be characterized, synthesizing demographic studies between Roman's wide-separation planets and transit and radial velocity surveys' small-separation planets will provide a new window into \etaearth across the Galaxy.

\subsection{Astrometry}\label{astrometry}

At its core, the astrometric technique determines the true positions of objects --- and it is from measuring how these positions change that we detect orbiting companions such as planets or stars. Informative reviews of this detection technique can be found in \citet{Quirrenbach2010} and \citet{Perryman2018}. The sensitivity of the technique thus depends on:
\begin{enumerate}
    \item the physical system properties, which dictate the companion's effect on the host star's motion, and 

    \item instrumental and observational effects, which must be corrected to reveal the objects' true motions.
\end{enumerate}
Astrometry's potential contribution to constraining \etaearth depends on how well these factors can be characterized. 

It is difficult to generalize the limits imposed by the instrumental/observational factors because they depend wholly on the observer's instruments and access to innovative techniques. These elements include, for example, the position-measuring technique (``centroiding''), the surface shapes of the observer's CCD and filter(s), and (if the observer is Earth-bound) the refractive effect of the atmosphere. These factors can affect a target's apparent position by up to multiple milliarcseconds (mas), which is similar --- and often larger --- than the signal induced by small companions (discussed below).

For the physical system properties that affect astrometry's sensitivity, the general rule is that a target will have more apparent motion induced by a companion if the system is more nearby or if its orbital semi-major axis is larger, as long as the star's companion does not contribute significant light to the system total.
This caveat is necessary because a star-star system's apparent semi-major axis depends also on its light ratio\footnote{The ``target'' for which we measure motions is the combined light (photocenter) of all unresolved objects in that system. For a star-planet pair, this combined light is centered on the primary star, but for a star-star system the center of light is \textit{between} those two objects, reducing the apparent size of the semi-major axis. Thus for any apparently small semi-major axis that could be a potential astrometric planet detection, a stellar companion must carefully be ruled out.}. For the purposes of this discussion, however, we will consider only star-planet systems in which the companion is dark.

\begin{figure}
\includegraphics[width=0.45\textwidth]{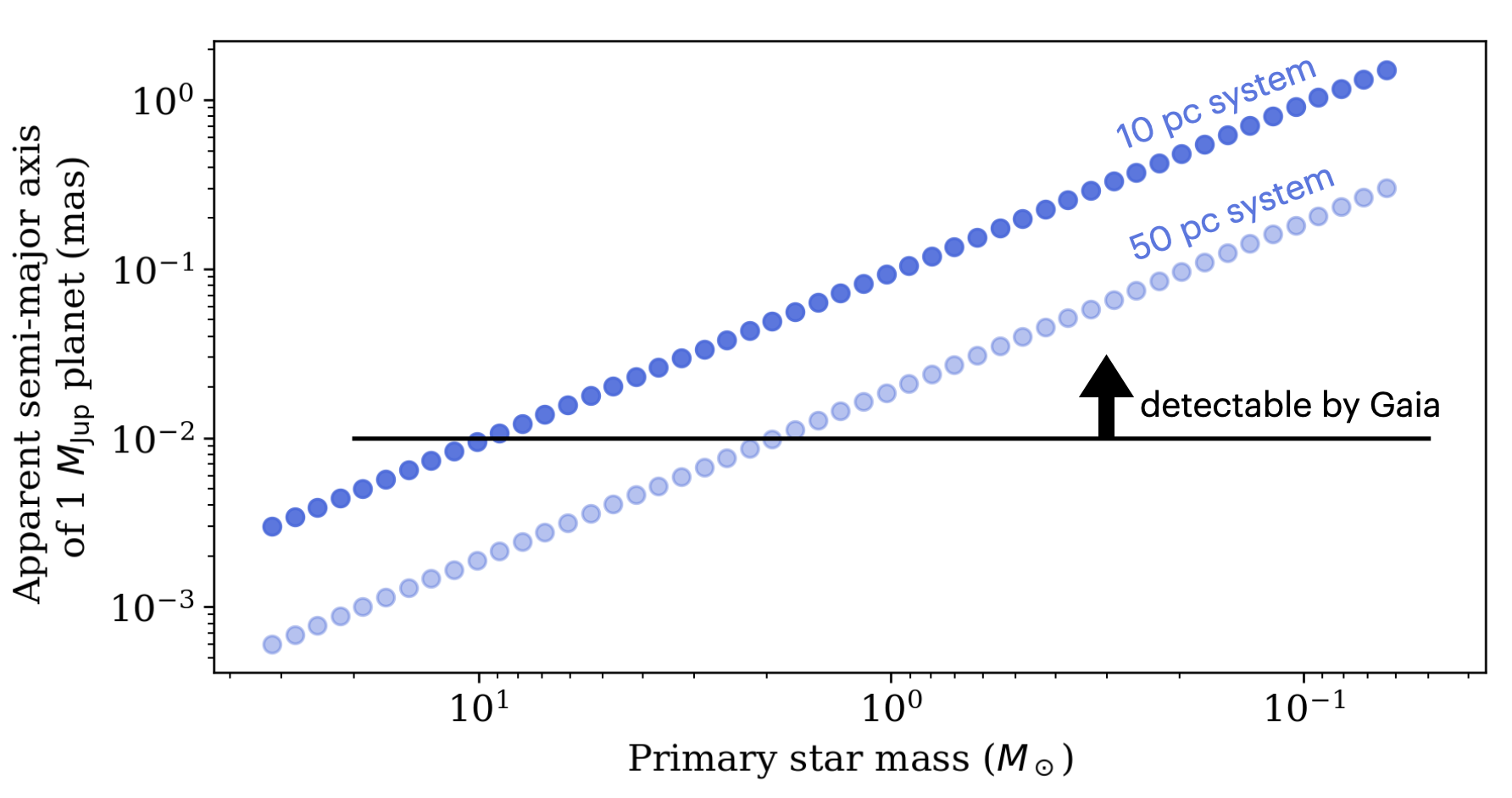} 
\caption{Size of astrometric orbit for a 1~\Mjup companion as a function of its host star mass. The hypothetical systems in dark blue are at distances of 10~pc, and those in light blue are at 50~pc. A 1~\Mearth planet would induce a semi-major axis on its host star roughly 300 times smaller --- i.e., below the \gaia detection limit here for all host star masses.
\label{fig:astrometry}}
\end{figure}

The astrometric signature of a planet with mass $M_{\rm p}$ and semi-major axis $a$ orbiting a star with mass $M_\star$ that is a distance $d$ parsecs away is given by 
\begin{equation}
    \theta = 3 \, \mu {\rm as} \, \left( \frac{a}{\rm au} \right)  \left( \frac{M_{\rm p}}{M_\oplus} \right) \left( \frac{M_\star}{M_\odot} \right)^{-1} \left( \frac{d}{\rm pc} \right)^{-1}.
\end{equation}
Thus, for a Sun-like star with an Earth-mass companion at 1~au, that star's orbit around its center of mass is only $3 \times 10^{-6}$~au --- translating to 0.003~mas at 1~pc, which is far below the threshold of recent ground- and space-based efforts (0.010--10~mas; see below). The size of this signal decreases linearly with distance, but fortunately increases linearly as stellar host mass decreases: in the above example, replacing the Sun-like star with an M dwarf of 0.3~\Msun yields a 0.01~mas orbit. Figure~\ref{fig:astrometry} demonstrates this relationship. Although this is still too small to be traced by current efforts, it suggests that low-mass stars are a promising regime in which astrometry can contribute to \etaearth.

Recent efforts in astrometry include both ground- and space-based programs, but none to date have been able to detect Earth-mass companions around Sun-like stars. For ground-based astrometry, the longest-running effort is that of the REsearch Consortium On Nearby Stars \citep[RECONS;][]{Henry2018}, which has been ongoing since 1999 and tracks M dwarf stars, achieving sensitivity of $\sim$10~mas \citep[corresponding to 10--30~\Mjup for some systems, and as low as $\sim$1~\Mjup for others;][]{Lurie2014}. 
The Carnegie Astrometric Planet Search \citep[CAPS;][]{Boss2009} ran for 12~years and targeted M, L, and T dwarfs, also achieving nearly Jupiter-mass sensitivity \citep{Boss2017,Boss2019}. The decade-long Hawaii Infrared Parallax Program \citep{Dupuy2012} monitored several dozen M, L, and T dwarfs, combining their data with Keck AO to determine masses for most of these \citep{Dupuy2017}, with the least massive at $\sim$30~\Mjup. Finally, the Stellar Planet Survey \citep[STEPS;][]{Pravdo2003STEPS,Pravdo2005STEPS,Pravdo2009STEPS} began a survey in 1998 of 30 M dwarfs in search of giant planets, reporting candidate companions down to giant planet masses and ruling out multi-\Mjup planets in several cases. 

Using the different ground-based approach of very-long-baseline interferometry, \citet{GRAVITY2024} identified a candidate 2.1~$M_\mathrm{Neptune}$ planet orbiting one of the two M dwarfs in the GJ~65 binary. By combining the light from multiple telescopes separated by tens of meters, the GRAVITY instrument on the VLTI \citep[Very Large Telescope Interferometer;][]{GRAVITY2017} stands out among ground-based techniques with its impressive sensitivity to sub-Jupiter mass planets around lower-mass stars.

Among space-based efforts, the \gaia mission has already detected several dozen planets \citep{Gaia2023DR3summary, Gaia2023DR3planets}, all firmly Jupiter-mass or greater. Future \gaia data releases are expected to increase this planet count a hundred- or thousand-fold, fully fleshing out the Jupiter-mass regime but with no effective sensitivity to Earth-mass planets
\citep{Perryman2014}; \gaia's sensitivity (visualized in Figure~\ref{fig:astrometry}), is limited to $\gtrsim$10~microarcsec by most optimistic estimates. 
Additionally, several groups are using the \gaia results in conjunction with Hipparcos data to detect and characterize low-mass companions with better precision than ground-based data alone \citep[e.g.][]{Brandt2019,Feng2022,Li2023,Barbato2024}. However, these detections to date are all at least brown dwarf or Jupiter-mass. 

Astrometry holds additional promise when combined with radial velocity efforts. 
This era has already begun: RV measurements have recently confirmed \gaia detections of one 12~\Mjup planet and one 21~\Mjup brown dwarf \citep{Stefansson2025}.
In a similar vein, the planned effort by \citet{Yahalomi2023} would combine ground-based HARPS-3 RVs from the Terra Hunting Experiment \citep[THE;][]{Hall2018} with space-based astrometry from \gaia and the Roman Space Telescope \citep{Spergel2015_Roman}. The inclusion of astrometric data promises to break the degeneracy between mass and inclination measured by the radial velocity, improving the sensitivity to Earth-like planets around Sun-like stars by a factor of $\sim$10. To this end, THE will observe ~$\sim$40 solar-type stars for 10 years, with the potential to detect Earth-, Saturn-, and Jupiter-mass planets.

Several future efforts show promise toward finding Earth-mass planets through innovative astrometric techniques. 
The Telescope for Orbit Locus Interferometric Monitoring of Our Astronomical Neighborhood \citep[TOLIMAN][]{TOLIMAN2021,TOLIMAN2024_sim,TOLIMAN2024} is a mission concept that would search for planets orbiting in the HZ of nearby Sun-like stars in binary systems, using these bright sources to avoid some typical challenges of astrometric measurements (e.g., having a bright companion as a reference star will significantly boost the astrometric signal-to-noise).\footnote{The mission concept has recently been renamed to SHERA (Searching for Habitable Exoplanets with Relative Astrometry; \citet{Christiansen2025}).} 
The proposed Closeby Habitable Exoplanet Survey \citep[CHES;][]{CHES2022,CHES2024a,CHES2024b} would search for Earth-mass planets in the HZ of Sun-like stars via a spacecraft at the Sun-Earth L2 point. 
The Theia satellite mission concept \citep{Theia2017,Malbet2022} would be capable of finding habitable-zone Earths and super-Earths around the nearest Sun-like stars by taking a high-precision ``point-and-stare'' approach.
And NASA's Roman Space Telescope, which will cover the sky for several years at a variety of cadences, has potential to find Earth-mass planets, especially in combination with earlier astrometry data from \gaia \citep{Melchior2018,WFIRSTAstrometryWorkingGroup2019}. 

\section{Summary \& Conclusions}
\label{sec:conclusions}

A key component informing mission yield predictions and identifying target stars is the precise determination of the frequency of Earth-like planets in the HZ of Sun-like stars, or \etaearth. The quantitative assessment of \etaearth is contingent on several distinct stellar and planetary parameters and analytical techniques, some of which require focused attention in order to contribute towards \etaearth, illustrated in Figure \ref{fig:etaEarthOverview}.

As the number of planets increased and the definitions of \etaearth converged, the values of \etaearth varied widely (Figure \ref{fig:etaEarthHistory}), reflecting the variations between analyses in the treatments, catalogs, stellar properties, and the result of extrapolations due to a very low number of detections in the \etaearth size and period regime (Section \ref{sec:etaEarthHistory}). 
Theoretical studies for better informing extrapolations as well as future observations would help with the convergence of \etaearth values. Several considerations impact \etaearth measurements; the key factors identified so far have been summarized in Section \ref{sec:missingAspects}. It is important to reiterate that this list is not exhaustive. Defining habitability and identifying the factors that influence it are multifaceted questions, underscoring the vast scope for studies on planet evolution and habitability. The parameters that contribute towards calculation of \etaearth are:
\begin{itemize}
    \item \textbf{Stellar Multiplicity: } 
    Stellar multiplicity has become easier to detect with adaptive optics and \gaia, but binary stars remain a source of contamination for single-star samples (Figure \ref{fig:sullivan2022}). Studies of occurrence rates of planets in binaries are needed to determine whether demographic properties like \etaearth are the same in single and multi-star systems.
    \item \textbf{Planet Multiplicity: } 
    The \kepler survey revealed a plethora of multi-planet systems, with $\sim40\%$ of the transiting planets belonging to multi-transiting systems and an even higher fraction for the rate of intrinsic multi-planet systems. Yet, so far, no studies have considered the potential impact of planet multiplicity on \etaearth, due to the model degeneracies between multiplicity and mutual inclinations, and the need to extrapolate to the HZ of Sun-like stars. Future work is necessary to model planet multiplicity in \etaearth calculations, and may enable a distinction between the overall occurrence vs. the fraction of stars with HZ Earth-like planets.
    \item \textbf{Conditional Occurrence Rate of \etaearth: } The paucity of observed systems with true Earth analogs makes it difficult to study how the presence of these planets could correlate with others (e.g., Jupiters; Figure \ref{fig:RosenthalSample}). More data are needed to refine the Earth-Jupiter correlation or improve models to the point of extrapolation.
    \item \textbf{Chemical Composition: } 
    The environment in which planets form impacts the likelihood for rocky planet formation, architecture, and habitable zone. Stellar abundances will be a critical input for \etaearth calculations, as they are a proxy for the natal disk and impact the bounds of the habitable zone.
    \item \textbf{Age: } 
    Atmospheric mass loss and inward migration shape the observed population of small, short-period planets, which \kepler relies on to estimate \etaearth. Yet, because these processes operate differently, or not at all, at larger separations, extrapolating short-period demographics into the habitable zone introduces significant uncertainties (Figure \ref{fig:occ_comp}). Distinguishing stripped sub-Neptunes from “born rocky” planets is essential for making reliable inferences about Earth-like planets in the habitable zone.
    \item \textbf{Galactic Population: } 
    Recent work suggests that stellar clustering, galactic velocity, and environmental perturbation related to host star galactic motion may affect rocky planet occurrence and habitability (Figure \ref{fig:zink_galactic_occurrence}); however, this relationship is not yet well constrained for planets of any type.
    \item \textbf{Spectral Type: } Limited sample sizes and large uncertainties currently obscure whether the occurrence-stellar mass trends observed for close-in planets also extend to the habitable zone across stellar spectral types (Figure \ref{fig:eta_spty}). In addition, the large difference in the habitable zone distance for M dwarfs adds an additional factor to consider (Figure \ref{fig:HZ_bounds}).
\end{itemize}

One recurring requirement for better understanding and constraining the parameters influencing \etaearth discussed in Section \ref{sec:missingAspects} is the dearth of detections of Earth-like planets and the imperative to extend the sample size of known planets to Earth-like planets. Many observing techniques have been used to attempt to constrain \etaearth, with various levels of precision (discussed in Section \ref{sec:futureProspects}). Inconsistencies between  different methods and low detection sample size have contributed to the highly uncertain values currently predicted. Additional sensitivity from future telescopes and mission concepts will extend the requisite parameter space. Section \ref{sec:futureProspects} reviews where our detection techniques currently stand in their ability to resolve earth-sized planets in the HZ of sun-like stars.
\begin{itemize}
    \item \textbf{Transits: } 
    Transits surveys, in particular the \kepler mission, are responsible for our current best estimates of \etaearth, though these estimates are still limited by too few detections and require extrapolation from planets with shorter orbital periods and larger radii. However, ongoing efforts with TESS seem promising to improve estimates of \etaearth for M dwarfs, and several upcoming missions, are poised to detect $\sim$dozens of Earth-sized planet candidates in the HZ of both low-mass and Sun-like stars by increasing sample sizes (e.g., Roman, Earth 2.0) and focusing on bright stars with better photometric precision (e.g. Plato), giving every indication that these new data will significantly constrain \etaearth by the mid 2030s. 
    \item \textbf{Radial Velocity: } 
    Existing RV data are insufficient to measure \etaearth without extrapolating from occurrence rates of more massive planets (Figure \ref{fig:rv}). However, substantial effort continues to be put towards furthering the planet detection capabilities of RV observations, and optimistic predictions of future survey yields show that they may soon enable a robust measurement of \etaearth. 
    \item \textbf{Imaging: } Current constraints on $\eta_{\oplus}$ from direct imaging are highly uncertain (consistent with 100\%) due to the lack of detections. In the near future ($\sim$10 yrs), ELTs will enable constraining $\eta_{\oplus}$ to $\leq$35\% from the seven closest FGK stars (assuming the null result, or no detections). In the more distant (decades) future, purpose-built observatories such as HWO observing hundreds of FGK stars will enable constraining $\eta_{\oplus}$ to within a few percent, with an upper limit of $\sim$1\% being possible in the event of no detections.
    \item \textbf{Microlensing: } 
    Microlensing has its highest sensitivity to habitable zone planets near the galactic center, a different stellar population than the solar neighborhood. Interpolation of monolithic surveys between microlensing (Roman) and transiting (\kepler, Roman) planet samples should provide a stronger constraint on \etaearth than extrapolations of either alone.
    \item \textbf{Astrometry: } 
    Recent and ongoing observational efforts, both ground- and space-based, have been sensitive to companions only Jovian-mass and greater, thus have yielded no direct information on \etaearth (Figure~\ref{fig:astrometry}). Future efforts in the coming decade have the potential to break through this mass sensitivity floor by combining astrometry with other techniques (e.g., RVs) and by implementing novel space-based observational approaches.
\end{itemize}

In the decades to follow, as we uncover previously unidentified factors influencing \etaearth and expand the known planet population across multiple parameter spaces, our current definitions and inferences related to \etaearth and habitability might undergo multiple revisions. Reaching a comprehensive, robust occurrence rate of Earth-like planets is contingent on the allocation of concerted resources to methodologies, studies, and technological advancements in detection techniques, such as those highlighted in this review. The search for life, or, more precisely, for biosignatures indicating evidence of Earth-like life, relies on this vital groundwork to quantitatively constrain \etaearth. There is much work still to be done in order to quantify, discover, and explore how unique or ubiquitous Earth analogs are. However, in research timescales, we are tantalizingly close, so long as we sustain, if not increase, our current momentum.

\section{Acknowledgments}

This review paper benefited from discussions held within the NASA Exoplanet Program Analysis Group (ExoPAG) and its Science Interest Group 2 (SIG 2), which focuses on exoplanet demographics. While not produced under the auspices of ExoPAG or SIG 2, the paper reflects insights informed by community engagement in those forums.

This project was supported in part by an appointment to the NRC Research Associateship Program at the U.S. Naval Research Laboratory in Washington, D.C., administered by the Fellowships Office of the National Academies of Sciences, Engineering, and Medicine. Basic Research in Astronomy at the U.S. Naval Research Laboratory is supported by 6.1 Base funding. 

Support for this work was provided by NASA through the NASA Hubble Fellowship grant HST-HF2-51497 awarded by the Space Telescope Science Institute, which is operated by the
Association of Universities for Research in Astronomy, Inc., for NASA, under contract NAS5-26555.

S.G. is supported by an NSF Astronomy and Astrophysics Postdoctoral Fellowship under award AST-2303922.

G.D.M. acknowledges support from FONDECYT project 1252141 and the ANID BASAL project FB210003.

M.Y.H. and A.C. acknowledge support from the NASA Postdoctoral Program at NASA Ames Research Center, administered by Oak Ridge Associated Universities under contract with NASA.

Part of this research was carried out at the Jet Propulsion Laboratory, California Institute of Technology, under a contract with the National Aeronautics and Space Administration.

E.H.V. acknowledges support from Five Colleges, Inc. as the Five College Astronomy Department Research/Education Postdoctoral Fellow.

M.K. acknowledges the support of the Natural Sciences and Engineering Research Council of Canada (NSERC), RGPIN-2024-06452. Cette recherche a été financée par le Conseil de recherches en sciences naturelles et en génie du Canada (CRSNG), RGPIN-2024-06452.

This work was partially supported by funding from the Center for Exoplanets and Habitable Worlds. The Center for Exoplanets and Habitable Worlds and the Penn State Extraterrestrial Intelligence Center are supported by Penn State and its Eberly College of Science.

Part of this research was carried out at the Jet Propulsion Laboratory, California Institute of Technology, under a contract with the National Aeronautics and Space Administration.

\clearpage
\bibliographystyle{aasjournalv7}
\bibliography{main}

\end{document}